\documentstyle[a4p,11pt,epsfig,array,cite,pennames,amssymb]{article}
\newcommand{\inmath}[1] {\ifmmode#1\else$#1$\fi}
\newcommand{\definmath}[2] {\def#1{\ifmmode#2\else$#2$\fi}}
\definmath{\PWpm} {\mathrm{W}^{\pm}}      
\definmath{\Pgtp} {\tau^{+}}        
\definmath{\Pgtm} {\tau^{-}}        
\definmath{\Pgtpm}   {\tau^{\pm}}         
\definmath{\Pgn}  {\nu}          
\definmath{\Pagn} {\overline{\nu}}     
\definmath{\Pq}      {\mathrm{q}}
\definmath{\Paq}  {\overline{\mathrm{q}}}
\definmath{\Pu}      {\mathrm{u}}
\definmath{\Pau}  {\overline{\mathrm{u}}}
\definmath{\Pd}      {\mathrm{d}}
\definmath{\Pad}  {\overline{\mathrm{d}}}
\definmath{\Ps}      {\mathrm{s}}
\definmath{\Pas}  {\overline{\mathrm{s}}}
\definmath{\Pc}      {\mathrm{c}}
\definmath{\Pac}  {\overline{\mathrm{c}}}
\definmath{\Pb}      {\mathrm{b}}
\definmath{\Pab}  {\overline{\mathrm{b}}}
\definmath{\Pt}      {\mathrm{t}}
\definmath{\Pat}  {\overline{\mathrm{t}}}
\definmath{\Pap}  {\overline{\mathrm{p}}}
\definmath{\Pan}  {\overline{\mathrm{n}}}
\definmath{\PaD}  {\overline{\mathrm{D}}}
\definmath{\PaDz} {\overline{\mathrm{D}}^{0}}
\definmath{\PaB}  {\overline{\mathrm{B}}}
\definmath{\PaBz} {\overline{\mathrm{B}}^{0}}
\definmath{\PsDpm}   {\mathrm{D}^{\pm}_{\mathrm{s}}}  
\definmath{\PcgLpm}  {\Lambda^{\pm}_{\mathrm{c}}}  
\definmath{\PD} {\mathrm{D}}     
\definmath{\PDst} {\mathrm{D}^{*}}     
\definmath{\PgLz} {\Lambda^{0}}        
\newcommand {\lsp}      {{{\tilde{\chi}}^{0}}_{1}}
\newcommand {\nln}      {{{\tilde{\chi}}^{0}}_{2}}
\newcommand {\Gravitino} {\tilde{\mathrm{G}}}
\newcommand {\nngg}       {\nu\overline{\nu}\gamma\gamma}
\newcommand {\nnggbra}       {\nu\overline{\nu}\gamma(\gamma)}
\newcommand {\nngggbra}       {\nu\overline{\nu}\gamma\gamma(\gamma)}
\newcommand {\ra}         {\rightarrow}
\newcommand {\ee}         {\mathrm{e}^+ \mathrm{e}^-}

\newcommand{\epem}   {\Pep\Pem}
\newcommand{\gamgam} {\Pgg\Pgg}
\newcommand{\mumu}   {\Pgmp\Pgmm}
\newcommand{\tautau} {\Pgtp\Pgtm}
\newcommand{\nunu}   {\Pgn\Pagn}

\newcommand{\eetogg}    {\epem\to\gamgam}

\newcommand{\eetomumu}     {\epem\to\mumu}
\newcommand{\eetotautau}   {\epem\to\tautau}

\newcommand{\eetonunu}     {\epem\to\nunu}

\newcommand{\BR}             {{\mathrm BR}}
\newcommand{\PX}             {{\mathrm X}}
\newcommand{\PY}             {{\mathrm{Y}}}
\newcommand{\eetonngg}       {\epem \to \nngg}
\newcommand{\eetonnggbra}    {\epem \to \nnggbra}
\newcommand{\eetonngggbra}   {\epem \to \nngggbra}
\newcommand{\eetoXY}         {\epem \to \PX\PY}
\newcommand{\eetoXX}     {\epem \to \PX\PX}
\newcommand{\XtoYg}       {\PX \to \PY\gamma}

\newcommand{\sigbrXY}    {\sigma(\eetoXY)\cdot\BR(\XtoYg)} 
\newcommand{\sigbrXX}    {\sigma(\eetoXX)\cdot\BR^2(\XtoYg)} 
\newcommand{\eetoXXs}   {\epem \to \nln \nln}
\newcommand{\eetoXYs}   {\epem \to \nln \lsp}
\newcommand{\XtoYgs}    {\nln \to \lsp\gamma}
\newcommand{\XtoYgg}    {\lsp \to \Gravitino\gamma}
\newcommand{\mx}         {M_{\PX}}
\newcommand{\my}         {M_{\PY}}
\newcommand{\myzero}        {\my\approx 0}

\newcommand{\mxmax}      {\mx^{\rm max}}

\newcommand{\roots} {\sqrt{s}}

\newcommand{\costhe} {\cos\theta}
\newcommand{\acosthe}   {|\costhe\,|}

\definmath{\GeV}  {\mathrm{GeV}}
\definmath{\GeVc} {\mathrm{GeV}\!/c}
\definmath{\GeVcc}   {\mathrm{GeV}\!/c^2}
\definmath{\MeV}  {\mathrm{MeV}}
\definmath{\MeVc} {\mathrm{MeV}\!/c}
\definmath{\MeVcc}   {\mathrm{MeV}\!/c^2}
\definmath{\MVm}  {\mathrm{MV}\!/\mathrm{m}}
\definmath{\keV}  {\mathrm{keV}}
\definmath{\keVcm}   {\mathrm{keV}\!/\mathrm{cm}}
\definmath{\kV}      {\mathrm{kV}}
\definmath{\km}      {\mathrm{km}}
\definmath{\meter}   {\mathrm{m}}
\definmath{\cm}      {\mathrm{cm}}
\definmath{\mm}      {\mathrm{mm}}
\definmath{\micron}  {\mu\mathrm{m}}
\definmath{\nm}      {\mathrm{nm}}
\definmath{\kg}      {\mathrm{kg}}
\definmath{\gram} {\mathrm{g}}
\definmath{\second}  {\mathrm{s}}
\definmath{\microsec}   {\mu\mathrm{s}}
\definmath{\degree}  {^\circ}
\definmath{\degC} {^\circ\mathrm{C}}
\definmath{\ohm}  {\Omega}
\definmath{\Mohm} {\mathrm{M}\Omega}
\definmath{\rad}  {\mathrm{rad}}
\definmath{\mrad} {\mathrm{mrad}}
\definmath{\nb}      {\mathrm{nb}}

\newcommand{\eqref}[1]  {(\ref{#1})}
\newcommand{\NIM} {Nucl.~Instrum.\ Methods}

\newcommand{\OPALColl}  {OPAL Collab.}

\newcolumntype{L} {>{$}l<{$}}
\newcolumntype{C} {>{$}c<{$}}
\newcolumntype{R} {>{$}r<{$}}

\begin{document}
\begin{titlepage}
\begin{center}{\large   EUROPEAN LABORATORY FOR PARTICLE PHYSICS
}\end{center}
\begin{flushright}
      CERN-EP/98-143 \\
08$^{\mathrm{th}}$ September 1998
\end{flushright}
\bigskip\bigskip\bigskip
\boldmath
\begin{center}{\huge\bf  Search for Anomalous Photonic \\
Events with Missing Energy in $\epem$ Collisions \\
\vspace{1mm}
 at $\roots$ = 130, 136 and 183~$\GeV$
}\end{center}
\unboldmath
\bigskip\bigskip
\begin{center}{\LARGE The OPAL Collaboration
}\end{center}
\bigskip

\begin{abstract} 
Photonic events with large missing energy have been observed in 
$\mathrm e^+e^-$ collisions at centre-of-mass energies of 130, 136 
and 183~GeV collected in 1997
using the OPAL detector at LEP. Results are presented 
for event topologies with a single photon 
and missing transverse energy or with an acoplanar photon pair.
Cross-section measurements are performed within
the kinematic acceptance of each selection. These results are compared with the
expectations from the Standard Model process $\eetonunu$ + photon(s).
No evidence is observed for 
new physics contributions to these final states. 
Using the data at $\roots$ = 183 GeV,
upper limits on $\sigbrXY$ and $\sigbrXX$ are derived for the case
of stable and invisible $\PY$. These limits apply to 
single and pair production of excited 
neutrinos ($\PX = \nu^*, \PY = \nu$), to neutralino production ($\PX=\nln, \PY=\lsp$)
and to supersymmetric models in which $\PX = \lsp$ and $\PY=\Gravitino$ 
is a light gravitino.

\end{abstract}


\vspace{2.0cm}
\begin{center}
(Submitted to Eur. Phys. J. {\bf C}.)
\end{center}

\vspace{1cm}

\vspace{5mm}
\end{titlepage} 
\begin{center}{\Large        The OPAL Collaboration
}\end{center}\bigskip
\begin{center}{
G.\thinspace Abbiendi$^{  2}$,
K.\thinspace Ackerstaff$^{  8}$,
G.\thinspace Alexander$^{ 23}$,
J.\thinspace Allison$^{ 16}$,
N.\thinspace Altekamp$^{  5}$,
K.J.\thinspace Anderson$^{  9}$,
S.\thinspace Anderson$^{ 12}$,
S.\thinspace Arcelli$^{ 17}$,
S.\thinspace Asai$^{ 24}$,
S.F.\thinspace Ashby$^{  1}$
% ...
}\end{center}\bigskip
\bigskip
$^{  1}$School of Physics and Astronomy, University of Birmingham,
Birmingham B15 2TT, UK
\newline
$^{  2}$Dipartimento di Fisica dell' Universit\`a di Bologna and INFN,
I-40126 Bologna, Italy
\newline
$^{  3}$Physikalisches Institut, Universit\"at Bonn,
D-53115 Bonn, Germany
\newline
$^{  4}$Department of Physics, University of California,
Riverside CA 92521, USA
\newline
$^{  5}$Cavendish Laboratory, Cambridge CB3 0HE, UK
\newline
$^{  6}$Ottawa-Carleton Institute for Physics,
Department of Physics, Carleton University,
Ottawa, Ontario K1S 5B6, Canada
\newline
$^{  7}$Centre for Research in Particle Physics,
Carleton University, Ottawa, Ontario K1S 5B6, Canada
\newline
$^{  8}$CERN, European Organisation for Particle Physics,
CH-1211 Geneva 23, Switzerland
\newline
$^{  9}$Enrico Fermi Institute and Department of Physics,
University of Chicago, Chicago IL 60637, USA
\newline
$^{ 10}$Fakult\"at f\"ur Physik, Albert Ludwigs Universit\"at,
D-79104 Freiburg, Germany
\newline
$^{ 11}$Physikalisches Institut, Universit\"at
Heidelberg, D-69120 Heidelberg, Germany
\newline
$^{ 12}$Indiana University, Department of Physics,
Swain Hall West 117, Bloomington IN 47405, USA
\newline
$^{ 13}$Queen Mary and Westfield College, University of London,
London E1 4NS, UK
\newline
$^{ 14}$Technische Hochschule Aachen, III Physikalisches Institut,
Sommerfeldstrasse 26-28, D-52056 Aachen, Germany
\newline
$^{ 15}$University College London, London WC1E 6BT, UK
\newline
$^{ 16}$Department of Physics, Schuster Laboratory, The University,
Manchester M13 9PL, UK
\newline
$^{ 17}$Department of Physics, University of Maryland,
College Park, MD 20742, USA
\newline
$^{ 18}$Laboratoire de Physique Nucl\'eaire, Universit\'e de Montr\'eal,
Montr\'eal, Quebec H3C 3J7, Canada
\newline
$^{ 19}$University of Oregon, Department of Physics, Eugene
OR 97403, USA
\newline
$^{ 20}$CLRC Rutherford Appleton Laboratory, Chilton,
Didcot, Oxfordshire OX11 0QX, UK
\newline
$^{ 22}$Department of Physics, Technion-Israel Institute of
Technology, Haifa 32000, Israel
\newline
$^{ 23}$Department of Physics and Astronomy, Tel Aviv University,
Tel Aviv 69978, Israel
\newline
$^{ 24}$International Centre for Elementary Particle Physics and
Department of Physics, University of Tokyo, Tokyo 113-0033, and
Kobe University, Kobe 657-8501, Japan
\newline
$^{ 25}$Institute of Physical and Environmental Sciences,
Brunel University, Uxbridge, Middlesex UB8 3PH, UK
\newline
$^{ 26}$Particle Physics Department, Weizmann Institute of Science,
Rehovot 76100, Israel
\newline
$^{ 27}$Universit\"at Hamburg/DESY, II Institut f\"ur Experimental
Physik, Notkestrasse 85, D-22607 Hamburg, Germany
\newline
$^{ 28}$University of Victoria, Department of Physics, P O Box 3055,
Victoria BC V8W 3P6, Canada
\newline
$^{ 29}$University of British Columbia, Department of Physics,
Vancouver BC V6T 1Z1, Canada
\newline
$^{ 30}$University of Alberta,  Department of Physics,
Edmonton AB T6G 2J1, Canada
\newline
$^{ 31}$Duke University, Dept of Physics,
Durham, NC 27708-0305, USA
\newline
$^{ 32}$Research Institute for Particle and Nuclear Physics,
H-1525 Budapest, P O  Box 49, Hungary
\newline
$^{ 33}$Institute of Nuclear Research,
H-4001 Debrecen, P O  Box 51, Hungary
\newline
$^{ 34}$Ludwigs-Maximilians-Universit\"at M\"unchen,
Sektion Physik, Am Coulombwall 1, D-85748 Garching, Germany
\newline
\bigskip\newline
$^{  a}$ and at TRIUMF, Vancouver, Canada V6T 2A3
\newline
$^{  b}$ and Royal Society University Research Fellow
\newline
$^{  c}$ and Institute of Nuclear Research, Debrecen, Hungary
\newline
$^{  d}$ and Department of Experimental Physics, Lajos Kossuth
University, Debrecen, Hungary
\newline
$^{  e}$ on leave of absence from the University of Freiburg
\newline

\clearpage\newpage
\section{ Introduction }
\label{sec:intro}
 
We describe measurements and searches using
a data sample of
photonic events with large missing
energy collected in 1997 with the OPAL detector at LEP.
The events result
from $\epem$ collisions at centre-of-mass energies 
of 130.0, 136.0 and 182.7~GeV with integrated luminosities of 
2.35, 3.37 and 54.5~pb$^{-1}$, respectively. 
The present paper complements our recent publication
of results from earlier data samples \cite{OPALSP172} 
consisting of a total of 25~pb$^{-1}$ at centre-of-mass energies
of 130, 136, 161 and 172~GeV. 
Results on photonic events without missing energy 
at $\sqrt{s}=$ 183~GeV are presented 
in a separate paper\cite{OPALgg183}.
Recent measurements of 
photonic event production have also been made by the other LEP
collaborations at centre-of-mass energies above the W pair 
threshold~\cite{LEP2SP}, including new results at $\sqrt{s}=$ 
183~GeV~\cite{ALEPHSP183}.
Related searches in $\rm{p}\overline{{\rm{p}}}$ collisions
have been reported in~\cite{Tevatron}.
The interest in the 1997 data is twofold. The main 
motivation is that the large  
data set at a higher centre-of-mass energy (183 GeV) 
gives discovery potential in a new kinematic regime. 
Additionally, the lower energy data sets allow the experiment 
at 130 and 136 GeV to be repeated leading to
a check of our previous results which 
indicated an excess of events with acoplanar photons~\cite{OPALSP172}. 

The single-photon  and acoplanar-photons search topologies presented here
are designed to select events with one or more photons and significant missing 
transverse energy, indicating  the presence of at least one neutrino-like 
invisible particle which interacts only weakly with matter.
The event selections for these search topologies
are similar to those used in our recent publication\cite{OPALSP172}.
The main changes are improvements in the 
rejection of backgrounds from cosmic ray interactions which allow for 
an increased kinematic acceptance of the single-photon topology.
The single-photon search topology is sensitive to neutral events in which 
there are one or two photons and missing energy which, within the 
Standard Model, are expected from the $\eetonnggbra$ 
process\footnote{The photon in parentheses denotes that the
presence of this photon is allowed but not required.}.
The allowance for two observed photons retains acceptance 
for doubly radiative neutrino pair production.
The acoplanar-photons search topology is designed to select neutral events with 
two or more photons and significant missing transverse energy
which, within the Standard Model, are expected 
from the $\eetonngggbra$ process. The selection is 
designed to retain acceptance for events with three photons, 
if the system formed by the three photons 
shows evidence for significant missing transverse energy.  

These photonic final-state topologies are sensitive to new physics of the 
type $\eetoXY$ and $\eetoXX$ where $\PX$ is neutral and decays radiatively 
($\XtoYg$) and $\PY$ is stable and only weakly interacting. For the general case of
massive $\PX$ and $\PY$ this includes conventional supersymmetric
processes\cite{Kane} $(\PX = \nln, \PY = \lsp)$. 
In this context it has been emphasized \cite{radN2} that the
radiative branching ratio of the $\nln$ may be large.
These topologies also have particularly good sensitivity
for the special case of $\myzero$.
This applies both to the production of
excited neutrinos $(\PX = \nu^*, \PY = \nu)$ and to supersymmetric models
in which the lightest supersymmetric particle (LSP) is a light 
gravitino\footnote{The mass scale is typically $\cal{O}$(keV).}
and $\lsp$ is the next-to-lightest supersymmetric particle (NLSP) which decays 
to a gravitino and a photon ($\PX=\lsp, \PY=\Gravitino$).
For supersymmetric models with a light gravitino,
the photonic branching ratio of the $\lsp$ is naturally large. 
Such a signature has been discussed in~\cite{ELLHAG} and more recently 
in~\cite{gravitinos,gravitinos2,LNZ,chang} in the context of both 
no-scale supergravity models and gauge-mediated supersymmetry breaking models.
Other types of new physics to which these search topologies are sensitive include
the production of an invisible particle in association with a photon
and the production of invisible particles tagged by initial-state radiation.
An example of this is $\epem \ra \Gravitino \Gravitino \gamma$\cite{Zwirner}.
The acoplanar-photons search topology also has sensitivity to the
production of two particles, one invisible, or with an invisible decay mode, 
and the other decaying into two photons. 
Such events might arise from the production of a 
Higgs-like scalar particle, $\rm S^0$ :
$\rm e^+e^-\rightarrow Z^0S^0$, followed by
S$^0$$\rightarrow \gamma\gamma$, $\rm Z^0\rightarrow \nu\overline{\nu}$.
OPAL results from a search for this process at $\roots = 183$ GeV, including the 
hadronic and leptonic $\rm Z^0$ decays, are reported in a
separate paper\cite{OPAL_Hgg183}.
The results from our previous searches~\cite{OPALSP172}
for $\eetoXY$ and $\eetoXX$ with $\XtoYg$
have been used to set model-dependent limits on excited 
neutrinos and neutralinos \cite{OPAL_CLIENTS}. The new
results reported here can be used in the same manner.

This paper will first briefly describe the detector
and the Monte Carlo samples used.
The event selection for each search topology will 
then be described, followed by cross-section measurements for 
$\eetonnggbra$ and $\eetonngggbra$ and comparisons with Standard Model 
expectations.
Implications of these results on the possibility of new physics processes
will be discussed. 

\section{Detector and Monte Carlo Samples}
\label{sec:detector}

The OPAL detector, which is described in detail in~\cite{OPAL-detector},
contains 
a pressurized
central tracking system operating inside
a solenoid with a magnetic field of 0.435 T.
The region
outside the 
solenoid (barrel) and the pressure bell (endcap) is 
instrumented with scintillation counters, presamplers and the 
lead-glass electromagnetic calorimeter (ECAL).
The magnet return yoke is instrumented for hadron calorimetry and
is surrounded by external muon chambers.
Electromagnetic calorimeters close to the beam
axis
measure luminosity and complete
the acceptance.

The measurements presented here are mainly based on the observation of 
clusters of energy deposited in the lead-glass electromagnetic 
calorimeter. This consists of an array of 9,440 lead-glass blocks
in the barrel ($|\cos{\theta}| < 0.82$) with a
quasi-pointing geometry and two dome-shaped endcap arrays, each of 1,132
lead-glass blocks,
covering the polar angle\footnote{In the OPAL coordinate system, 
$\theta$ is the polar angle defined with respect to the electron 
beam direction and $\phi$ is the azimuthal angle.} 
range ($0.81 < |\cos{\theta}| < 0.984$).
Energies measured in the ECAL 
will normally refer to those obtained after corrections to account for
losses in upstream material. Energies without these corrections  
are called deposited energies.
In the regions ($0.72 < \acosthe < 0.82$) and  
($\acosthe > 0.945$), the energy resolution of the ECAL is degraded 
relative to the nominal resolution. This degradation is largely
due to increased amounts of material in front of the ECAL.
In some cases (where stated) these 
regions have been excluded from the analysis. 
Fully hermetic electromagnetic calorimeter coverage is achieved 
beyond the end of the ECAL down to small polar angles with the use of the
the gamma-catcher calorimeter, the forward 
calorimeter (FD) and the silicon-tungsten calorimeter (SW). These
detectors cover the angular regions with respect to the beam of
140-205 mrad, 40-145 mrad and 24-59 mrad, respectively. 
However, a small region centred on a polar angle of 
30 mrad lacks useful calorimetric coverage due to the installation, in 1996, 
of a thick tungsten shield designed to protect the tracking chambers from 
synchrotron radiation background.
The effective limit of electromagnetic hermeticity is 
therefore around 33 mrad.

Scintillators in the barrel and endcap 
regions are used to 
reject backgrounds from cosmic ray 
interactions and to
provide time measurements for 
the large fraction ($\approx$ 80\%) of
photons which convert in the material in front of the ECAL.
The barrel time-of-flight (TOF) scintillator bars
are located outside the solenoid 
in front of the barrel ECAL
and match  
its geometrical 
acceptance ($|\cos{\theta}| < 0.82$).
Arrays of thin scintillating tiles with embedded 
wavelength-shifting-fibre readout \cite{OPAL-TE} 
were installed prior to the 1997 run.
The new tile endcap (TE) scintillator arrays are 
located at $0.81 < \acosthe < 0.955$
behind the pressure bell and 
in front of the endcap ECAL.

Additional scintillating tile arrays, referred to as the MIP-PLUG, 
were installed at polar angles between 40 and 200 mrad and consist of 
two pairs of scintillating tile layers designed for detection of muons.
The outer pair, covering the angular range of 125 to 200 mrad is used
in this paper to provide redundancy in the rejection of
events with energetic electromagnetic showers in the gamma-catcher region.

The tracking system, consisting 
of a silicon microvertex detector (SI), a vertex drift chamber (CV) 
and a large volume
jet drift chamber (CJ), is used to reject events  with prompt charged particles. 
The silicon microvertex detector consists of two concentric cylindrical layers of 
silicon microstrip arrays, each layer providing both an azimuthal and longitudinal 
(along the beam direction) coordinate measurement. The two-layer acceptance 
covers $|\cos{\theta}|<0.90$.

Beam related backgrounds and backgrounds 
arising from cosmic ray interactions 
are rejected using the scintillator timing 
measurements and information 
from the electromagnetic calorimeter
shower shape, 
the hadron calorimeter and the muon detectors. 
The integrated luminosities of the data samples 
are determined to better than 1\% from small-angle Bhabha
scattering events in the SW calorimeter.
Triggers\cite{trigger} based on electromagnetic energy deposits in 
either the barrel or endcap electromagnetic calorimeters, and also  
on a coincidence of energy in the barrel electromagnetic calorimeter
and a hit in the TOF system, lead to 
full trigger efficiency for photonic events passing the event selection
criteria described in the following section.

For the expected Standard Model signal process,  
$\eetonunu$ + photon(s), the Monte Carlo generator 
KORALZ~\cite{KORALZ}
was used. 
For other expected Standard Model processes, a number of different 
generators were used: RADCOR~\cite{RADCOR} for $\epem \to \gamma \gamma (\gamma)$;
BHWIDE~\cite{BHWIDE} and TEEGG~\cite{TEEGG} for $\epem \to \epem (\gamma)$;
grc4f~\cite{GRC4F} for $\ee \ra \ell^+ \ell^- \nu\bar{\nu} (\gamma)$;
and KORALZ for  $\eetomumu(\gamma)$ and $\eetotautau(\gamma)$.
The expected contributions from each of these Standard Model processes were evaluated
using a total equivalent integrated luminosity at least ten
times larger than the integrated 
luminosity of the data sample.

To simulate possible new physics processes of the type  $\eetoXY$ 
and $\eetoXX$ where $\PX$ decays to $\PY\gamma$ and 
$\PY$ escapes detection, a modified version of
the SUSYGEN~\cite{SUSYGEN} Monte Carlo 
generator was used to produce neutralino pair events of the type 
$\eetoXYs$ and  $\eetoXXs$, $\XtoYgs$,  
with isotropic angular
distributions for the production and decay of $\nln$ and 
including initial-state radiation.
Monte Carlo events were generated at 40 (for $\PX\PY$ production) and 42
(for $\PX\PX$ production) points in the kinematically accessible region of the 
($\mx$, $\my$) plane.
All the Monte Carlo samples described above were processed through the
OPAL detector simulation~\cite{GOPAL}.

\section{Photonic Event Selection}
\label{sec:selection}

This section describes the criteria for selecting single-photon and 
acoplanar-photons events.
The kinematic acceptance of each selection is defined in terms of the photon energy, 
$E_{\gamma}$, and the photon polar angle, $\theta$. In addition,  
the scaled energy, $x_{\gamma}$, is defined as
$E_{\gamma}/E_{\mathrm{beam}}$, and the scaled transverse
energy, $x_{T}$, as $x_{\gamma}\sin{\theta}$.

\begin{description}
\item[Single-Photon -]
One or two photons accompanied by invisible particle(s):
\begin{itemize}
\item At least one photon with  $x_{T} > 0.05$ and with
$15^{\circ}<\theta<165^{\circ}$ ($\acosthe < 0.966)$.
\end{itemize}
\item[Acoplanar-Photons -] 
Two or more photons accompanied by invisible particle(s):
\begin{itemize}
\item At least two photons, each with $x_{\gamma} > 0.05 $
and $15^{\circ}<\theta<165^{\circ}$, or one photon 
with $E_{\gamma} > 1.75$ GeV  and $\acosthe < 0.8$ and a second photon
with $E_{\gamma} > 1.75$ GeV and $15^{\circ}<\theta<165^{\circ}$.
\item The scaled transverse momentum of the two-photon system 
consisting of the two highest energy photons,  
$p_T^{\gamma\gamma}$, must satisfy 
$p_T^{\gamma\gamma}/E_{\rm beam} > 0.05$.
\end{itemize}
\end{description}

In each of the two cases, it is desirable to retain acceptance for events 
with additional photons, if the resulting photonic system is
still consistent with the presence of significant 
missing energy. This reduces the sensitivity of each measurement to the
modelling of higher-order contributions.
Consequently,
a large fraction of the kinematic acceptance of the acoplanar-photons
selection is also contained in the kinematic acceptance of the
single-photon 
selection.

\subsection{Single-Photon Event Selection}
\label{sec:g1_selection}

The single-photon selection criteria are similar 
to the previous OPAL analysis of photonic events with missing 
energy\cite{OPALSP172} but have increased acceptance for 
lower energy photons 
due to improved rejection and control of 
cosmic ray and beam related backgrounds.
The modifications also improve the efficiency for events
with two detected photons.

\begin{itemize}

\item{
{\bf Kinematic acceptance.}
Events must contain a primary
electromagnetic cluster (that with the highest deposited energy in 
the barrel or endcap calorimeters) in the region 
$15^\circ < \theta < 165^\circ$
($\acosthe < 0.966$) with  $x_{T} > 0.05$.
Events are considered to have more than one photon if additional
electromagnetic clusters are found in the barrel or endcap
calorimeter ($\acosthe < 0.984$) having deposited energy
exceeding 300~MeV.
}

\item{
{\bf Cluster quality.} The primary electromagnetic cluster, combined 
with any clusters contiguous with it, must be consistent with the cluster size and 
energy sharing of blocks for a photon coming from near the interaction
point. The cluster size varies in both azimuthal and polar angle
extent as a function of $\acosthe$. The cluster extent cuts
are parametrized in $\acosthe$ accordingly.
Events are rejected if the cluster energy exceeds the 
beam energy by more than three standard deviations. 
}   

\item{
{\bf Forward energy vetoes.}
Events are rejected if
the energy sum of gamma-catcher clusters in either end is
greater than 5~GeV. Events are also rejected if
either
layer of the outer 
MIP-PLUG 
shows evidence of an energetic shower (pulse-height
exceeding about ten minimum ionizing particle equivalents).
An event can also be vetoed based on the transverse momentum sum of
clusters measured in the forward
calorimeters FD and SW. Events are rejected
that have a transverse momentum sum exceeding 1 GeV and where the
azimuthal angle of the transverse momentum sum is
within $60^{\circ}$ in azimuth of the 
direction opposite the measured momentum of the photonic system.
A final complementary veto 
rejects events that have an energy sum exceeding 5 GeV
where the sum is over all clusters in the FD, SW and 
forward part of the endcap ECAL ($\acosthe>0.966$)
which are within $60^{\circ}$ in azimuth of the
direction opposite the measured momentum of the photonic system.
The directional nature of these last two vetoes removes events with 
forward going high-energy particles that can account for some or all of
the missing transverse momentum. However, it minimizes losses from random 
noise or accidental energy deposits in the forward detectors. 
It also allows for the presence of initial-state radiated photons 
in the forward region which are not back-to-back with the photon(s).}

\item{
{\bf Muon veto.}
Events are rejected if there are any muon track segments
reconstructed in the muon chambers or 
in the hadron calorimeters.
Events are also rejected if there is significant
activity in the outer part of the barrel hadron calorimeter. 
The muon veto is used primarily to remove cosmic ray background.
}
\item{
{\bf Selective multi-photon veto.}
This veto addresses backgrounds, principally from 
$\Pep\Pem \to \Pgg\Pgg(\Pgg)$,
whilst retaining acceptance for events 
with two photons and missing energy. Events with a second
photon are rejected if any of the following criteria are satisfied:
\begin{itemize}
\item{The total energy of
the two clusters exceeds $0.9\roots$.}
\item{The acoplanarity angle\footnote{Defined as $180^{\circ}$ minus the 
opening angle in the transverse plane.}
of the two clusters is less than $2.5^\circ$.}
\item{The missing momentum vector 
calculated from the two clusters satisfies
$|\cos{\theta_{\mathrm{miss}}}| > 0.9$.}
\item{A third electromagnetic cluster is detected with
deposited energy exceeding 300~MeV.}
\item{$p_T^{\gamma\gamma}/E_{\rm beam} < 0.05$.}
\item{For events with at least one of the two clusters in the region
$\acosthe > 0.95$, the variable $b_{T}$ is less than
0.1, where $b_{T} = (\sin{\theta}_1 + \sin{\theta}_2) |\cos\left[(\phi_1 -
\phi_2)/2\right]|$. This amounts to a stronger acoplanarity cut for events
with at least one forward photon.}
\item{Events with an energy sum greater than 1 GeV
in either end of FD or SW are rejected if the 
two-photon plus forward photon system is planar, namely if
the sum of the three opening angles exceeds 350$^\circ$.}
\end{itemize}}

\end{itemize}

Events are required to either contain no reconstructed charged tracks
or contain a photon candidate consistent with a photon conversion
observed within the central tracking volume.
These events are referred to as non-conversion 
and conversion candidates, respectively. These two classes are mutually
exclusive. We will now describe in turn the additional criteria 
used for each candidate class.

{\bf Non-conversion candidates must satisfy the following 
additional criteria : }
\begin{itemize}
\item{
{\bf Charged track veto.}
It is required that there are no reconstructed tracks with 10 or more
hits in CJ.}
\item{
{\bf Timing requirements.}
The criteria
depend on the polar angle of the primary cluster.
For the angular region
$\acosthe<0.72$,
either an in-time associated TOF hit or the
absence of the special background vetoes (described
below) is required. The second requirement is made
in order to retain acceptance for
photons that do not convert before reaching the
TOF counters. 
For the range $0.72<\acosthe<0.82$, where the muon coverage
is not complete and the material in front of the ECAL leads to 
a high probability of a photon being detected in the TOF,
a good in-time
associated TOF hit is required. For $\acosthe>0.82$, a good in-time
associated TE hit is required as well as
the absence of the first three of the 
special background vetoes described below. 
A cluster with an associated TOF (TE) hit is considered to be in-time
if the measured arrival time of the photon at the TOF (TE)
is within 5 (30) ns  
of the expected time for a photon originating from the interaction point.
These definitions of in-time hits also apply to the single-photon
conversion selection as well as the acoplanar-photons selection. 
Events with a photon candidate having an out-of-time associated TOF or
TE hit are rejected as cosmic rays. 
}
\item{
{\bf Special background vetoes.}
Three special background vetoes are used for 
candidates that are in the endcap region or that have no TOF timing
information. A fourth special background veto is used only for 
photon candidates with $\acosthe <0.72$ that have no
TOF timing information.
The first veto rejects events in which any of the three
muon triggers\cite{trigger} 
(barrel and two endcaps) were present. This veto
rejects cosmic ray background. The second looks for a series of 
electromagnetic or hadronic
calorimeter clusters consistent with the same radial and azimuthal
position as the primary cluster, 
but at different positions along the beam direction. This veto
rejects beam halo backgrounds. The third looks for
a series of hits in the outer layers of the hadron calorimeter.
This veto rejects both cosmic rays and beam related backgrounds.
The fourth veto is based on 
the shape of the cluster
in the barrel ECAL.
The observed energy deposition in each 
lead-glass block of the cluster is fitted 
to the expected shower profile for a photon
coming from the interaction point.
One then calculates the 
variable S (in radiation lengths), 
defined as the difference between the measured 
and expected values of the energy-weighted lateral distance 
from the fitted shower centroid.
Events with S exceeding 0.2
are rejected.
This veto rejects  cosmic rays and beam related backgrounds,
both of which tend to have shower shapes wider than those 
of photons originating 
from the interaction point.
}
\end{itemize}

{\bf Conversion candidates must satisfy the following 
additional criteria : }
\begin{itemize}
\item{
{\bf Photon conversion consistency.} 
There must be at least one reconstructed charged track in the
central tracking chambers.
The charged track with the most hits
must be associated in space
to at least one of the two most energetic photon candidates.
In particular, the measured polar angle of the 
track should be consistent with the polar angle
of the ECAL cluster to within 100 mrad, and the 
azimuth at the point of closest approach of the 
track to the interaction point should differ by 
less than 100 mrad from the measured azimuthal angle of
the ECAL cluster. 
The azimuthal matching criterion is relaxed to 500 mrad for
clusters with $|\cos{\theta}|>0.90$ since 
conversions at forward angles often lead to large 
showers and difficulties in resolving the jet chamber 
left-right ambiguity.
In addition, for such forward photon conversion candidates, which mostly
convert in the CV endplate, it was further
required that there be at least 12 out of the first 16 wires hit in CJ.}

\item{
{\bf Prompt charged track veto.}
Events are rejected as being consistent with
containing a prompt charged track 
if at least one photon  
candidate
has azimuthally associated hits in the innermost
tracking detector
(SI for $\acosthe < 0.9$ and CV otherwise).
}

\item{
{\bf Two or more track veto.}
Events with conversion candidates are rejected if they have at
least two tracks, reconstructed from axial-wire hits in CV, 
with an opening angle in the transverse
plane exceeding 
$45^{\circ}$.
This criterion is used principally 
for redundancy in the rejection of Bhabha scattering events.}
\item{
{\bf Identified cosmic ray veto.}
Events with at least two electromagnetic clusters in the 
barrel region each with associated TOF hits are rejected
as identified cosmic rays if the time difference between the 
upper and lower TOF hits is consistent with a downward-going
cosmic ray.}
\item{
{\bf Timing requirements.}
The photon(s) associated with 
the track is required to have an in-time TOF or TE hit
depending on whether the polar angle of the photon matches
the TOF or TE geometrical acceptance. 
}
\end{itemize}

\begin{figure}[b]
\centerline{\epsfig{file=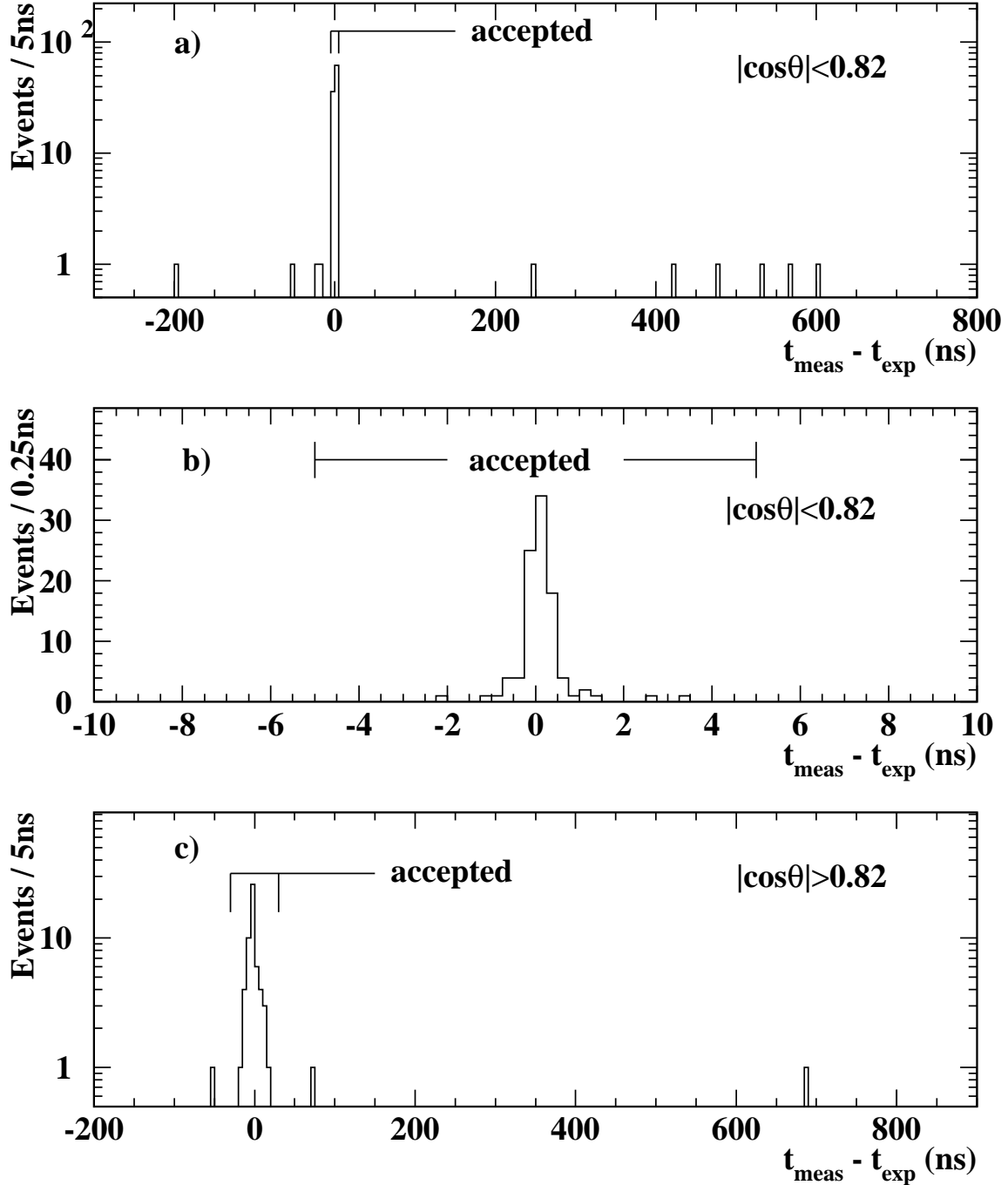,width=16cm,
bbllx=20pt,bblly=120pt,bburx=550pt,bbury=720pt}}
\caption{
For single-photon candidates in the data, a) shows the difference between 
the measured time-of-flight in the TOF and that 
expected for a photon from the interaction point for events
passing all cuts or failing only the TOF timing cut; b) the same plot
but magnifying the expected region for good events; c) shows
the difference between the measured time-of-flight in the TE
scintillators and that expected for a photon from the interaction
point for events passing all cuts or failing only the TE timing cut.
}
\label{f:cuts1}
\end{figure}

\begin{figure}[b]
\centerline{\epsfig{file=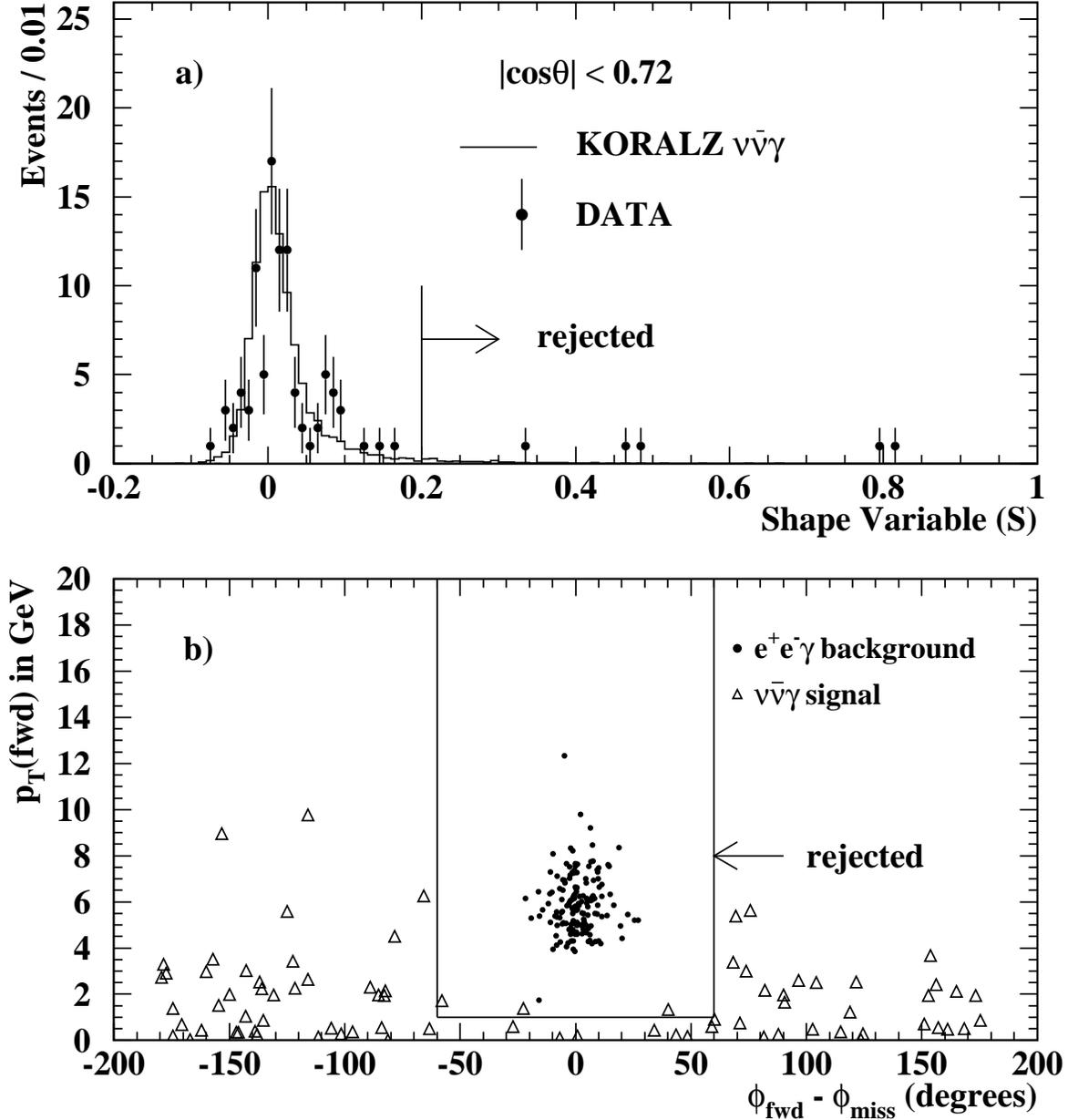,width=16cm,
bbllx=20pt,bblly=120pt,bburx=550pt,bbury=720pt}}
\caption{
a) The distribution for the cluster
shape variable S, for the KORALZ $\eetonunu \gamma(\gamma)$ Monte Carlo
(histogram) and for data (points). Events must have the primary
photon candidate within $\acosthe<0.72$ and passing all cuts or
failing only the special background vetoes. The Monte Carlo sample
is normalized to the integrated luminosity of the data.
b) The transverse momentum sum of hits measured in the 
FD and SW calorimeters is plotted against the
difference in azimuth between the forward momentum sum and the 
direction opposite the measured photonic system.
The events plotted are those passing all
cuts or failing only this cut.
The solid circles are from the 
$\epem \to \epem \gamma$ Monte Carlo while the open triangles are
from the $\eetonunu \gamma(\gamma)$ Monte Carlo. Most 
$\eetonunu \gamma(\gamma)$ events have
no significant forward energy and are not shown in the plot.
The number of events 
correspond to ten times the integrated luminosity of the data.
}
\label{f:cuts2}
\end{figure}

Distributions of some quantities used in the 
event selection are illustrated in Figures~\ref{f:cuts1} and \ref{f:cuts2}.
For photons with $\acosthe <0.82$, Figure~\ref{f:cuts1}a shows the 
difference between the measured TOF timing
and that expected for a photon 
originating from the interaction point for events passing
all selection criteria or failing only the TOF timing requirement. The eleven
events outside the accepted region of $\pm{5}$~ns are rejected as cosmic
rays. The expected region for good events is shown in greater detail in
Figure~\ref{f:cuts1}b.
Figure~\ref{f:cuts1}c shows the corresponding plot for
photons measured in TE with $\acosthe >0.82$. The three events 
outside the accepted region of $\pm{30}$~ns are rejected as cosmic rays.
Figure~\ref{f:cuts2}a shows the distribution of the 
cluster shape variable S for events passing
all selection criteria or failing only the cluster shape cut of the
special background vetoes. Rejecting events with S
greater than 0.2 
preserves the $\eetonnggbra$ Monte Carlo events,
while in the data it removes very large clusters 
which have been verified as being
due to cosmic rays and beam related backgrounds. 
Figure~\ref{f:cuts2}b shows the 
effect of one of the directional forward veto
cuts against its principal intended background, $\epem \to \epem \gamma$.
This cut is designed to remove events in
which a highly energetic particle travelling close to the beam
direction balances most or all of the transverse momentum
of the observed photon(s). The events plotted are those passing all
cuts or failing only this cut.
All the remaining $\epem \to \epem (\gamma)$ Monte Carlo
background events fall into this cut region while the signal 
from the $\eetonunu \gamma(\gamma)$ Monte Carlo has only a very small
number there. The data distribution for this plot was checked 
and found to agree well with the Monte Carlo prediction. 


\subsection{Acoplanar-Photons Event Selection}
\label{sec:g2_selection}

The acoplanar-photons selection has two overlapping regions of 
kinematic acceptance
in order to retain both sensitivity to low-energy photons and acceptance at large
$\acosthe$. These selections are based on analyses previously published by OPAL
using data collected at centre-of-mass energies of 130-172 GeV\cite{OPALSP172}.
Except where specified, cuts on photon candidates apply to the two highest
energy photon candidates found within the kinematic acceptance.  
The event selection criteria are described below:

\begin{itemize}
\item{
{\bf Kinematic acceptance.}
Events are accepted as candidates if there are at least two electromagnetic clusters
with scaled energy, $x_{\gamma}$, exceeding 0.05 in the polar angle region
$15^{\circ} < \theta < 165^{\circ}$ ($\acosthe < 0.966$). In order to retain 
sensitivity to physics processes producing low-energy photons, the minimum
energy requirement is relaxed to 1.5 GeV deposited energy
(corresponding to a photon energy of about 1.75~GeV\cite{OPALSP94}) for events
with a photon candidate in the polar angle region $\acosthe < 0.8$ provided
this photon is associated to an in-time TOF hit as outlined below
in the description of the timing requirements.
These two selections are referred to below as the ``high-energy'' and 
``low-energy'' selections, respectively. Background vetoes are applied
differently for the two parts of the selection, as described below.
The system consisting of the two highest energy photons 
must satisfy $p_T^{\gamma\gamma}/E_{\rm beam} > 0.05$.
}
\item{
{\bf Photon conversion consistency requirements or charged track veto.}
For the high-energy selection, events having
tracking information consistent with the presence of at least
one charged particle originating from the interaction point are rejected.
The rejection criteria are designed to retain acceptance for events in which 
one or both of the photons convert. 
Hit information from each of CJ, CV, 
and SI (for $\acosthe < 0.9$) are used
to form independent estimators for the existence of
charged particle activity. Events are rejected on the basis of
azimuthal association of charged particle activity
with the photon candidate clusters. 
To reject
$\rm e^+e^-\rightarrow\ell^+\ell^-\gamma\gamma$, an additional veto requires that 
there be no reconstructed charged track with transverse momentum exceeding 1~GeV,
with associated hits in CV,  
and separated from each of the photon candidates by more than $15^\circ$. 

The low-energy part of the selection does not allow photon conversions
in the tracking chambers. It requires that there be no 
reconstructed charged track in the event with 20 or more hits in CJ.
}
\item{
{\bf Cluster quality.} Photon candidates within the polar angle region 
$\acosthe < 0.75$ are required to have an angular cluster extent less
than 250 mrad in both $\theta$ and $\phi$. Additionally, to reduce background
from cosmic rays which graze the electromagnetic 
calorimeter producing extended
energy deposits that can be split by the clustering algorithm, photon candidates
are required to be separated by at least 2.5$^{\circ}$ in azimuth.
Events are rejected if a photon candidate cluster energy exceeds the 
beam energy by more than three standard deviations. 
}
\item{
{\bf Forward energy vetoes.}
The forward vetoes described for the single-photon selection
are applied with the same thresholds.
}
\item{
{\bf Muon veto.}
To suppress backgrounds arising from cosmic ray muon 
interactions or beam halo muons which can deposit significant 
energy in the calorimeter, the events must pass the muon veto 
described  for the single-photon selection. Additionally, the
first three individual vetoes of the 
special background vetoes described for the single-photon selection are 
applied to events in which no TOF information is present. 
}
\item{
{\bf Timing requirements.}
For the low-energy part of the selection, in order to ensure that the trigger is
fully efficient for low-energy photons,  we require that there be 
a photon in the 
barrel region with an associated in-time TOF hit.
For the high-energy selection
the event must have an associated in-time TOF or an associated in-time 
TE hit for at least one of the photon candidates. 
Events with a photon having an associated out-of-time TOF hit are rejected. 
Events in which one photon candidate has an associated out-of-time TE hit 
are retained provided they pass the other timing requirements.
Finally, if there is a charged track associated with a cluster within the polar 
angle region $\acosthe<0.82$, the requirement of an associated in-time TOF hit 
is applied. 
}

\item{
{\bf Selective multi-photon veto.}
As for the single-photon selection,
this veto is designed to reject backgrounds
primarily from $\Pep\Pem \to \Pgg\Pgg(\Pgg)$
whilst retaining acceptance for events 
with two or more photons and missing energy. Events are rejected
if any of the following criteria are satisfied:
\begin{itemize}
\item The total visible ECAL energy of the event exceeds $0.95\roots$.
\item The acoplanarity angle of the two highest energy 
clusters is less than $2.5^{\circ}$.
\item The missing momentum vector calculated from the two highest energy photon
candidates satisfies $\mathrm |\cos{{\theta}_{miss }}|~>~0.95$.
\item Events having three or more photon candidates 
(with deposited energy
greater than 300~MeV) 
are rejected unless the system formed by the three highest energy photons
is significantly aplanar 
(sum of the three opening angles $< 350^{\circ}$) and 
the transverse momentum of the three-photon system exceeds 0.1$E_{\rm beam}$.
For events with an energy sum greater than 1 GeV
in either end of FD or SW, the aplanarity cut is applied, using
the forward detector as the third photon candidate.
\end{itemize}
}
\end{itemize}

The acoplanar-photons selection described above has
a lower energy threshold for the most energetic photon than the single-photon
selection. However the single-photon selection has 
more acceptance for events without time-of-flight
information for the photons.
In order to obtain the best overall acceptance for acoplanar-photons, we 
have added to the above described acoplanar-photons selection
that part of the single-photon selection which is within the 
kinematic acceptance of the acoplanar-photons selection.
This addition results in a relative increase in
efficiency of 9\% for Standard Model $\eetonngggbra$ events.

\section{Results}
\label{sec:results}
The results of the single-photon 
and acoplanar-photons selections
are given below in sections~\ref{sec:sp_results}
and~\ref{sec:g2_results}.
The measured cross-sections for each search topology are given 
and compared with Standard Model expectations. As no
evidence for new physics processes is seen,
the results are presented in terms of
upper limits on
$\rm \sigma ( {\mathrm e^+e^-\rightarrow XY} ) 
\cdot BR ( {\mathrm X\rightarrow Y\gamma} )$
and 
\mbox{$\rm \sigma ( {\mathrm e^+e^-\rightarrow XX} ) 
\cdot BR^2 ( {\mathrm X\rightarrow Y\gamma} )$}.
This is done both 
for the general case of massive $\PX$ and $\PY$, applicable to conventional
supersymmetric models in which 
$\PX = \nln$ and  $\PY = \lsp$, 
and also separately for the special case of $\myzero$, which applies 
both to single and pair production of neutralinos 
in supersymmetric models in which the LSP is a light gravitino and to
single and pair production of excited neutrinos.
All efficiencies are evaluated with
the decay length of $\PX$ set to zero. 
%

For the purposes of new physics searches, only the 
$\roots = 183$ GeV data are considered; the $\roots = 130$ and 136 GeV
data do not open any new kinematic regions, nor is the integrated
luminosity at these energies sufficient to significantly improve the 
potential for discovery. 
For both the XX and XY searches, Monte Carlo samples
were generated for a variety of mass 
points in the kinematically accessible region of the $(\mx,\my)$ plane.
To set limits for arbitrary $\mx$ and $\my$,
the efficiency over the entire $(\mx,\my)$ plane is parametrized using the 
efficiencies calculated at the generated mass points.
In the single-photon search topology,
the regions with $\mx+\my < M_{\rm Z}$ 
are kinematically accessible at $\sqrt{s} \approx M_{{\rm Z}}$, 
and strong limits have already been 
reported\cite{LEP1XY}.
In the acoplanar-photons search topology, limits have been reported
for masses $\mx < M_{\rm Z}/2$\cite{LEP1XX}. In these low mass regions,
possible radiative return to the $\PZz$ followed by $\PZz\to$ XY or XX
would yield very different event kinematics than those produced by
the signal Monte Carlo generator.  
For these reasons, 
the search for XY production is restricted to 
the region with $\mx+\my > M_{\rm Z}$, and 
the search for XX production is restricted
to $\mx$ values larger than about $M_{\rm Z}$/2.

\boldmath
\subsection{Single-Photon}
\unboldmath
\label{sec:sp_results}

After applying the selection criteria of the single-photon
selection to the $\roots =$ 130, 136 and 183 GeV data samples,
a total of 21, 39 and 191 events are selected. 
The expected contributions from cosmic ray and beam related 
backgrounds are 0.02, 0.02 and 0.4 events, respectively. 
These backgrounds have been estimated from events
having out-of-time TOF or TE information  but passing all other selection
criteria and from events selected with looser criteria that have 
been visually scanned. 
Of the expected physics backgrounds from plausible sources, only
$\eetogg(\gamma)$, $\ee \ra \ell^+ \ell^- \nu\bar{\nu} (\gamma)$, 
$\ee \ra \mumu \gamma$ and
$\ee \ra \tautau \gamma$ have non-negligible contributions. Coming 
primarily from the $\ee \ra \ell^+ \ell^- \nu\bar{\nu} (\gamma)$
and $\mumu \gamma$ final states, the total physics
backgrounds contribute 0.07, 0.09 and 0.4 events, respectively, to
the 130, 136 and 183 GeV samples. The background contributions 
are summarized in Table~\ref{tab:sp_background}.  
For each of the three centre-of-mass energies,
Table~\ref{tab:sp_results} shows the number of events observed, 
the number of events expected from the Standard Model process 
$\eetonunu \gamma(\gamma)$ evaluated using
the KORALZ generator and 
the sum of background events expected from other Standard
Model physics processes with those from cosmic ray and beam related processes.
The numbers of events observed
agree with the numbers expected from $\eetonnggbra$
plus the background. 
The estimated efficiencies
for selecting $\eetonnggbra$ events within the
kinematic acceptance of the single-photon selection
are also given in Table~\ref{tab:sp_results}, as are the corresponding 
measured $\eetonnggbra$ cross-sections
within this kinematic acceptance, corrected for detector and selection
efficiencies, and subtracting the estimated background.
For both the single-photon and acoplanar-photons selections,
efficiency losses due to vetoes on random
detector occupancy range from about 
(2-4)\% at the different centre-of-mass energies. 
Quoted efficiencies include these losses.

\begin{figure}[b]
\centerline{\epsfig{file=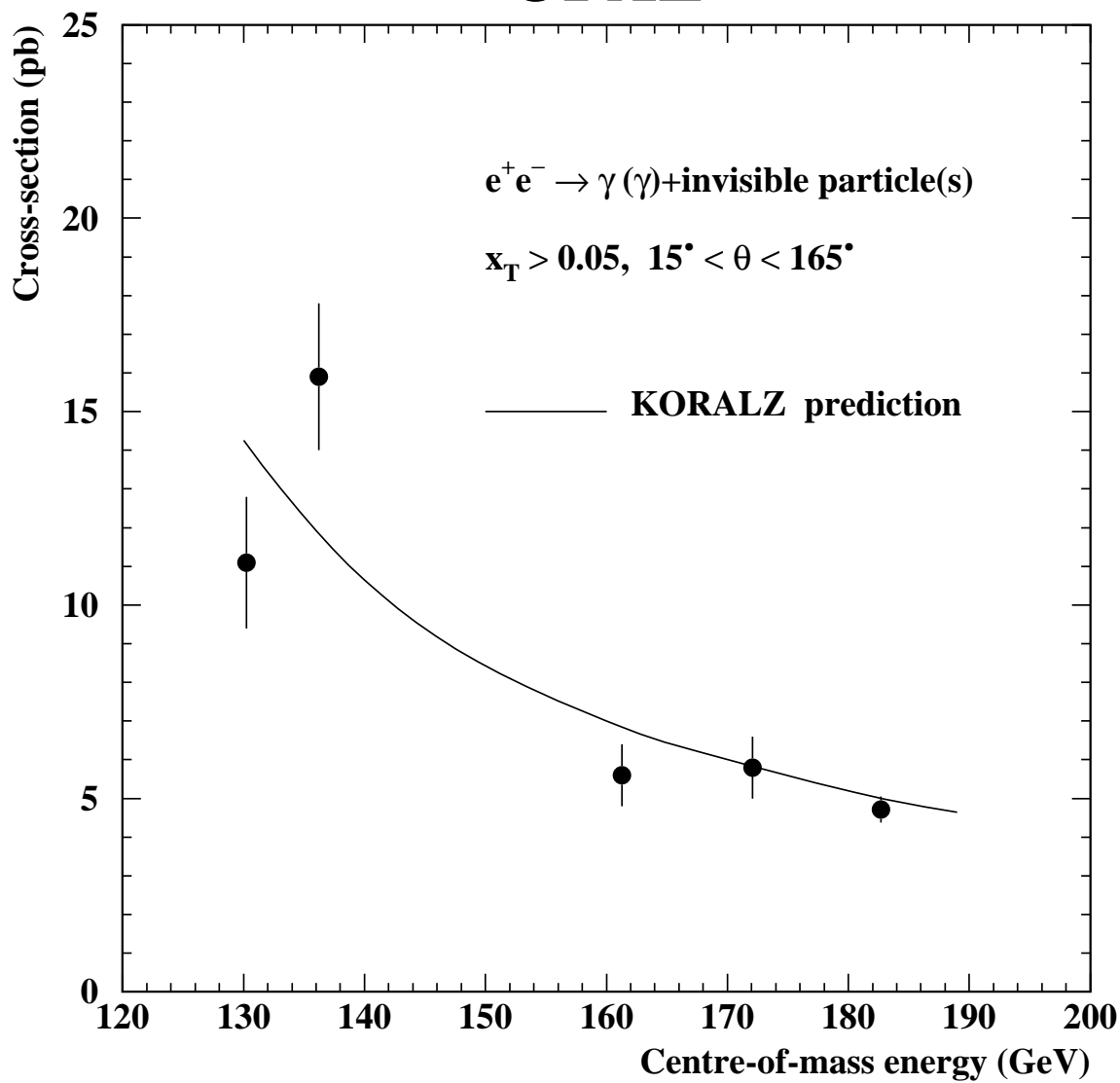,
width=16cm,bbllx=10pt,bblly=140pt,bburx=550pt,bbury=685pt}}
\caption{ The measured value of 
$\sigma(\epem \to \gamma (\gamma)$ + invisible particle(s)),
within the kinematic acceptance of the single-photon selection,
as a function of $\roots$.
The data points with error bars are OPAL
measurements at $\roots$ = 130, 136, 161, 172 and 183~GeV. 
The curve is the prediction for the
Standard Model process $\eetonnggbra$ from the KORALZ generator.
The data points at 130 and 136~GeV represent the 
weighted means of cross-section
measurements obtained from the 1995 and 1997 data samples.
The cross-section measurements from the previous data sets at
$\roots$ = 130, 136, 161 and 172~GeV have been corrected for the
difference in kinematic acceptance from the present analysis.
}
\label{f:sp_xs130_183}
\end{figure}

The total systematic error on the cross-section measurement
is estimated to be 3.5\%.
The contributing uncertainties 
are
from the integrated luminosity (0.5\%), 
effects due to uncertainties on the
electromagnetic calorimeter energy scale and
resolution (0.7\%) and the detector occupancy estimate (1\%).
In addition, the overall selection efficiency uncertainty (2\%)
is caused mainly by uncertainties in 
the simulation of the detector material and consequent 
photon conversion probabilities.
An additional error of 2.5\% is assigned 
based on the comparison of the estimated efficiency using 
two different event generators\cite{OPALSP172}.
The cross-section as a function of centre-of-mass
energy is plotted in Figure~\ref{f:sp_xs130_183}.
In this plot the measured cross-sections 
for $\sqrt{s}=130$ and 136 GeV are 11.1 $\pm$ 1.7 and 15.9 $\pm$ 1.9
pb, respectively. These
are weighted averages of the 1995 data results with the results
from this analysis. The results
for
$\sqrt{s}=161$ GeV
and $\sqrt{s}=172$ GeV are also plotted. 
The cross-section results from 
our earlier publication\cite{OPALSP172}
have been corrected for the slightly different kinematic
acceptance used in the analysis of those data. 
The corrected
cross-sections, obtained for 
the 1995 and 1996 data at $\sqrt{s}=$ 130, 136, 161 and 172 GeV are
10.6 $\pm$ 2.4, 17.3 $\pm$ 3.0, 5.6 $\pm$ 0.8 and 5.8 $\pm$ 0.8 pb,
respectively. 
For 1997 data at $\sqrt{s}=$ 130, 136 and 183 GeV the measured
cross-sections are 11.6 $\pm$ 2.5, 14.9 $\pm$ 2.4 and 
4.71 $\pm$ 0.34 pb, respectively.
The curve shows the
predicted cross-section from the KORALZ event generator
for the Standard Model process $\eetonnggbra$. 
The data are in reasonable agreement with the prediction.

\begin{figure}[b]
\centerline{\epsfig{file=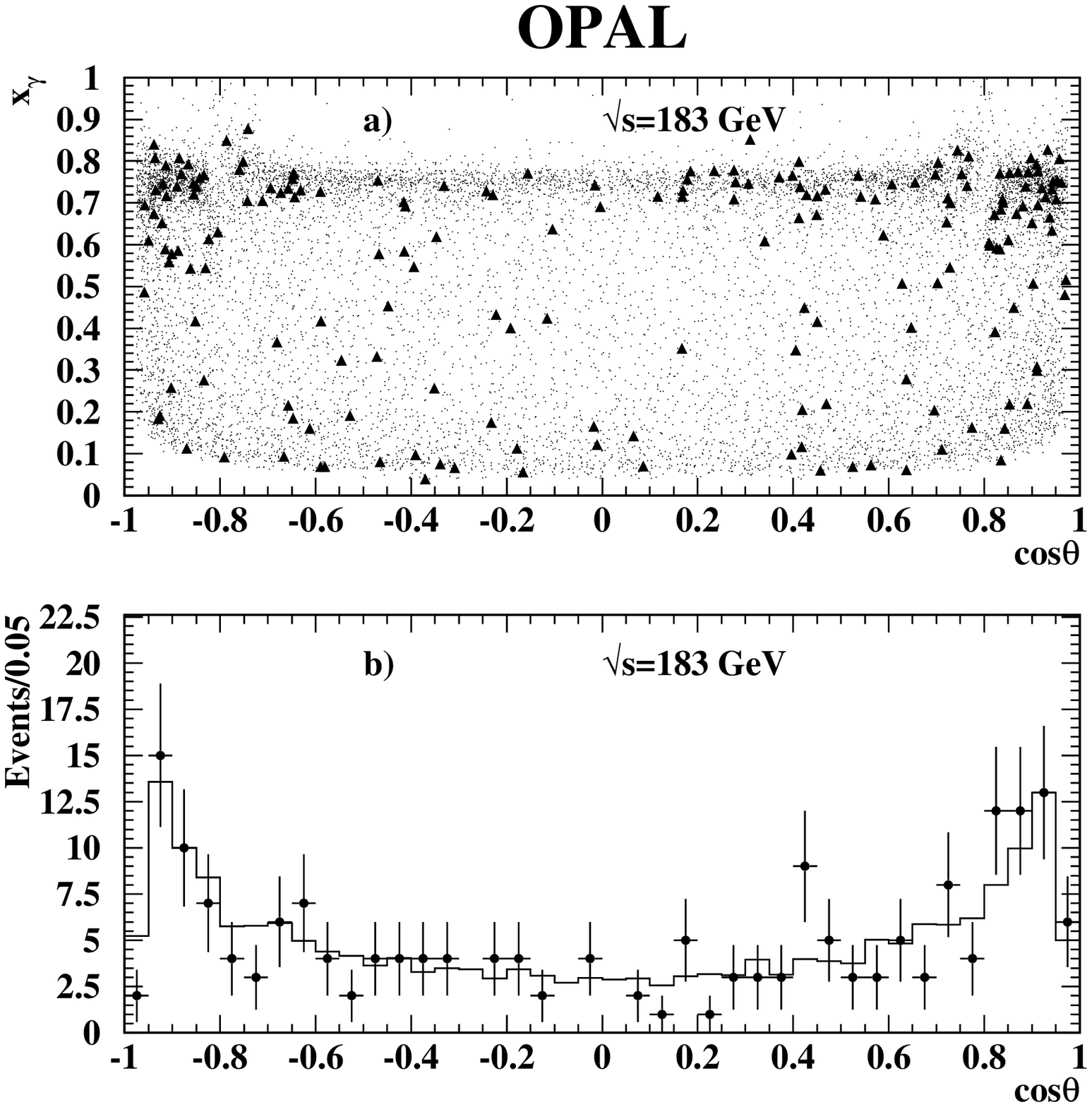,
width=16cm,bbllx=10pt,bblly=140pt,bburx=540pt,bbury=700pt}}
\caption{ a)~Distribution of $x_{\gamma}$ vs $\cos\theta$
for the most energetic photon in the single-photon selection
at $\roots$ = 183 GeV. 
The fine points are the KORALZ $\eetonnggbra$ Monte Carlo
(arbitrary normalization) and the solid triangles are the data.
b)~The $\cos\theta$ distribution for 
the most energetic photon in the single-photon selection
at $\roots$ = 183 GeV. 
The points with error bars are the data and the histogram is the
expectation from the KORALZ $\eetonnggbra$ Monte Carlo
normalized to the integrated luminosity of the data.
} 
\label{f:sp_xcosth183}
\end{figure}

\begin{figure}[b]
\centerline{\epsfig{file=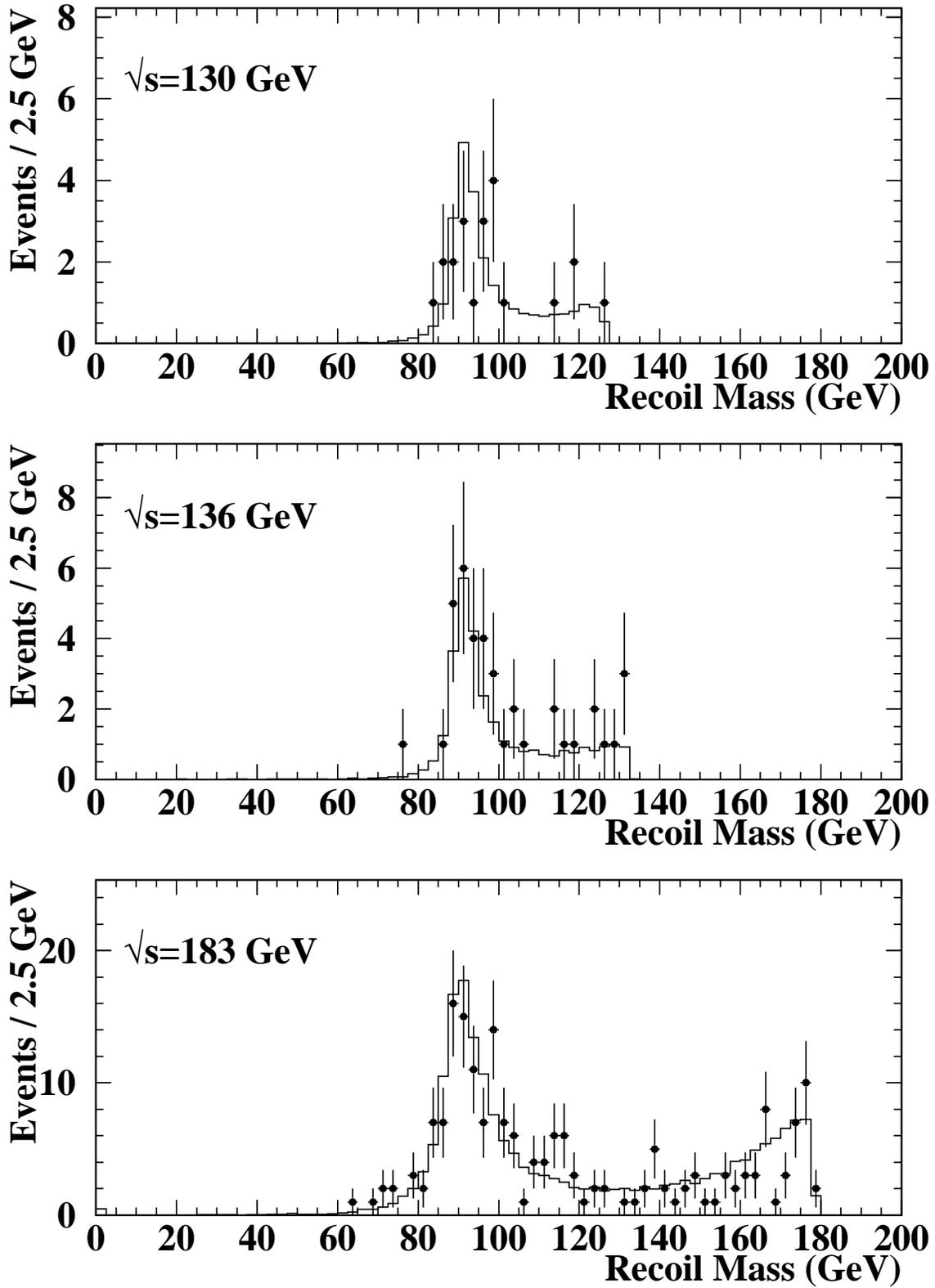,
height=22cm,bbllx=20pt,bblly=30pt,bburx=550pt,bbury=780pt}}
\caption{ The recoil mass distribution for events passing the
single-photon selection for the $\roots$ = 130, 136  
and 183 GeV data samples. 
The points with error bars are the data and the histograms are the
expectations from the KORALZ $\eetonnggbra$ Monte Carlo
normalized to the integrated luminosity of the data. 
}
\label{f:sp_mrec130_183}
\end{figure}

In Figure~\ref{f:sp_xcosth183}a, the scaled energy 
of the most energetic photon is plotted against the cosine
of its polar angle for events in the $\roots$=183 GeV sample.
The data are distributed as expected from the
$\eetonnggbra$ Monte Carlo. Similar agreement is found for
the 130 and 136 GeV data. In Figure~\ref{f:sp_xcosth183}b the 
polar angle distribution for the $\roots$=183 GeV
sample is shown and agrees with
the $\eetonnggbra$ Monte Carlo expectation.
If one calculates the recoil mass $M_{\rm recoil}$, defined as
the mass recoiling against the photon
(or against the two-photon system),
one expects a peak in the $M_{\rm recoil}$ 
distribution at $M_{\rm Z}$,
due to a large contribution from the decay $\PZz \to \nunu$.
One clearly sees this feature in the data as shown in 
Figure~\ref{f:sp_mrec130_183}.
There is good agreement between data 
and Monte Carlo in this distribution for each of the three centre-of-mass
energies.

The single-photon selection was designed to allow for the presence
of a second photon in order to accept events from the $\eetonngg$
process. In the $\roots$ = 130, 136 and 183~GeV data sets, 
2, 4 and 12 observed events are considered to
be two-photon events (i.e. have a second photon with deposited
energy exceeding 300~MeV in the ECAL). 
This is consistent with the expectations  
of 1.4, 1.7 and 11.3 events, respectively, from the KORALZ Monte Carlo.

\boldmath
\subsubsection{Search for $\eetoXY$, $\XtoYg$ ; General case: $\my\ge 0$}
\unboldmath
\label{sec:sp_results_allmy}

The single-photon selection described in Section~\ref{sec:selection} is
designed to maximize acceptance for Standard Model $\eetonnggbra$ events.
However, when searching for signatures of the process $\eetoXY$,~$\XtoYg$,
it is possible to implement further cuts to reduce the 
contribution from $\eetonnggbra$, which is
now considered as background.  
Depending on the values of $\mx$ and $\my$, various combinations of
the following cuts are applied to events in the single-photon sample:

\begin{itemize}
  \item {\bf Kinematic consistency:}  The energy of the most energetic
        photon is required to lie within the range kinematically
        accessible to a photon from the process $\eetoXY$, $\XtoYg$,
        after accounting for energy resolution effects.
  \item {\bf Degraded resolution:} Events with $M_{\rm recoil} < $75~GeV 
        are rejected if the most energetic photon lies in the region 
        $0.72 < |\cos\theta| < 0.82$ or $|\cos\theta| > 0.945$. 
        Energy resolution in these angular regions
        is known to be degraded.
  \item {\bf Low recoil mass:}  Require $M_{\rm recoil} < $75~GeV.
  \item {\bf Z$^0$ radiative return:}  The event is rejected if 
         $75~{\rm GeV} < M_{\rm recoil} < 105$~GeV.
\end{itemize}

The kinematic consistency cut and the degraded resolution cut
are applied to all $(\mx,\my)$ values.  
The low recoil mass cut is applied if $\my < 0.28\mx - 18$ (GeV), in which case
a significant portion of the expected photon energy distribution is
consistent with low recoil masses\footnote{This region was chosen so
as to optimize the expected sensitivity.  The optimization condition
chosen was that the expected upper limit on $\sigma\cdot \rm BR$ for new
physics contributions be minimized, 
where the expected upper limit is defined as the average limit one would expect to set in the 
absence of new physics contributions.
This definition has the advantage that it does not require one to specify 
the cross-section of possible new physics.}.  The radiative return cut is
applied if the low recoil mass cut is not applied, and if $\mx$ and $\my$ are 
such that the 
difference between the maximum and minimum kinematically allowed photon
energies (before energy resolution effects are considered) is greater
than $0.3 E_{\rm beam}$.

\begin{figure}[ht]
        \centerline{\epsffile{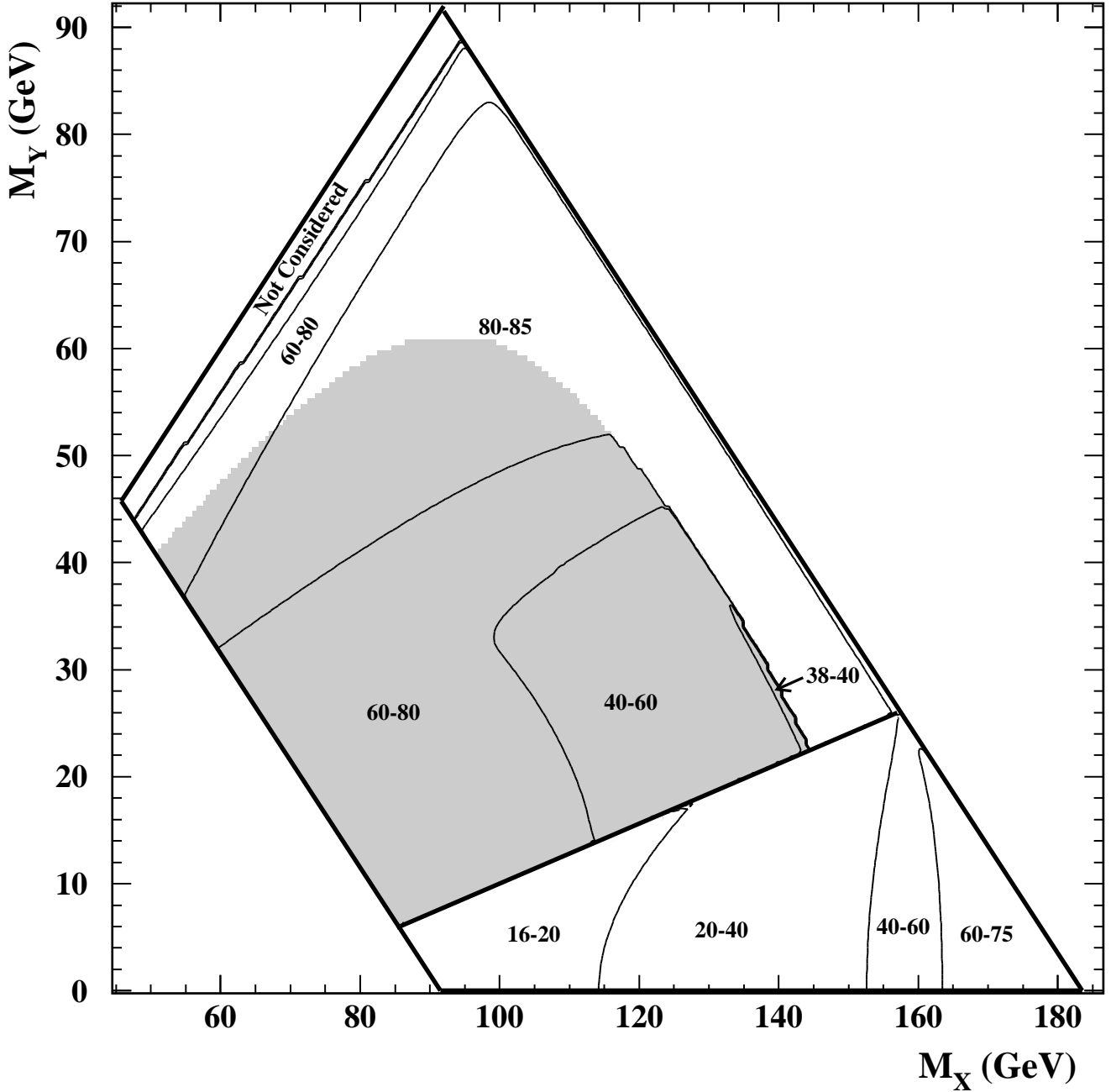}}
        \caption{ Single-photon selection efficiency contours (in \%)
                for $\sqrt{s} =$ 183 GeV as a function of $\mx$ and $\my$.
                Lines are drawn around the boundaries defined by 
                $\mx + \my = 183$ GeV, $\mx = \my$, and 
                $\mx + \my = M_{\rm Z}$, and to display the boundary  
                between the small and large $\my$ regions.  
		The shaded region indicates where
		the radiative return cut is applied. 
		The region in which $\mx - \my$ is small is not 
                considered, as explained in the text.}
        \label{f:sp_XYeff}
\end{figure}

The selection efficiencies including the above cuts as a 
function of $(\mx,\my)$ are 
given in Figure~\ref{f:sp_XYeff}. 
Also illustrated is the region in which the radiative return cut is applied.
In the region where $\mx - \my$ is
small, photon energies are correspondingly small and efficiencies 
become low.  
Since uncertainties due to energy scale and resolution effects
become significant in this region, we do not consider
values of $\mx$ and $\my$ that
lead to efficiencies of less than 40\% in the absence of the 
recoil mass or radiative return cuts.

\begin{figure}[ht]
        \centerline{\epsffile{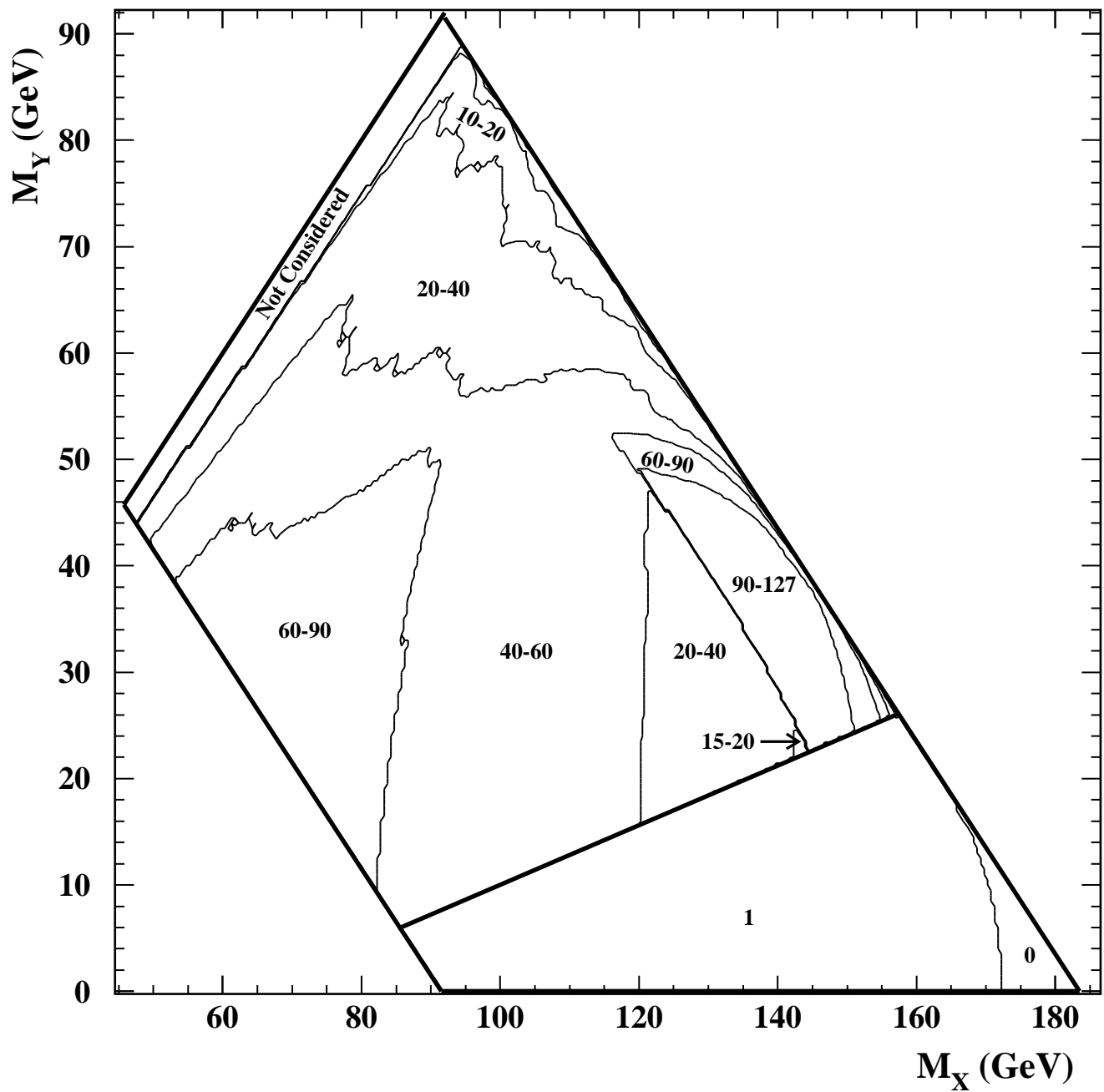}}
        \caption{Number of single-photon candidate events 
                in the $\sqrt{s} =$ 183 GeV data sample 
                consistent with each set of mass values ($\mx$, $\my$)
                for the process $\eetoXY$, $\XtoYg$.
                The boundaries and delineated regions are as defined for 
                Figure~\ref{f:sp_XYeff}.}
        \label{f:sp_evsXY_data}
\end{figure}

\begin{figure}[ht]
        \centerline{\epsffile{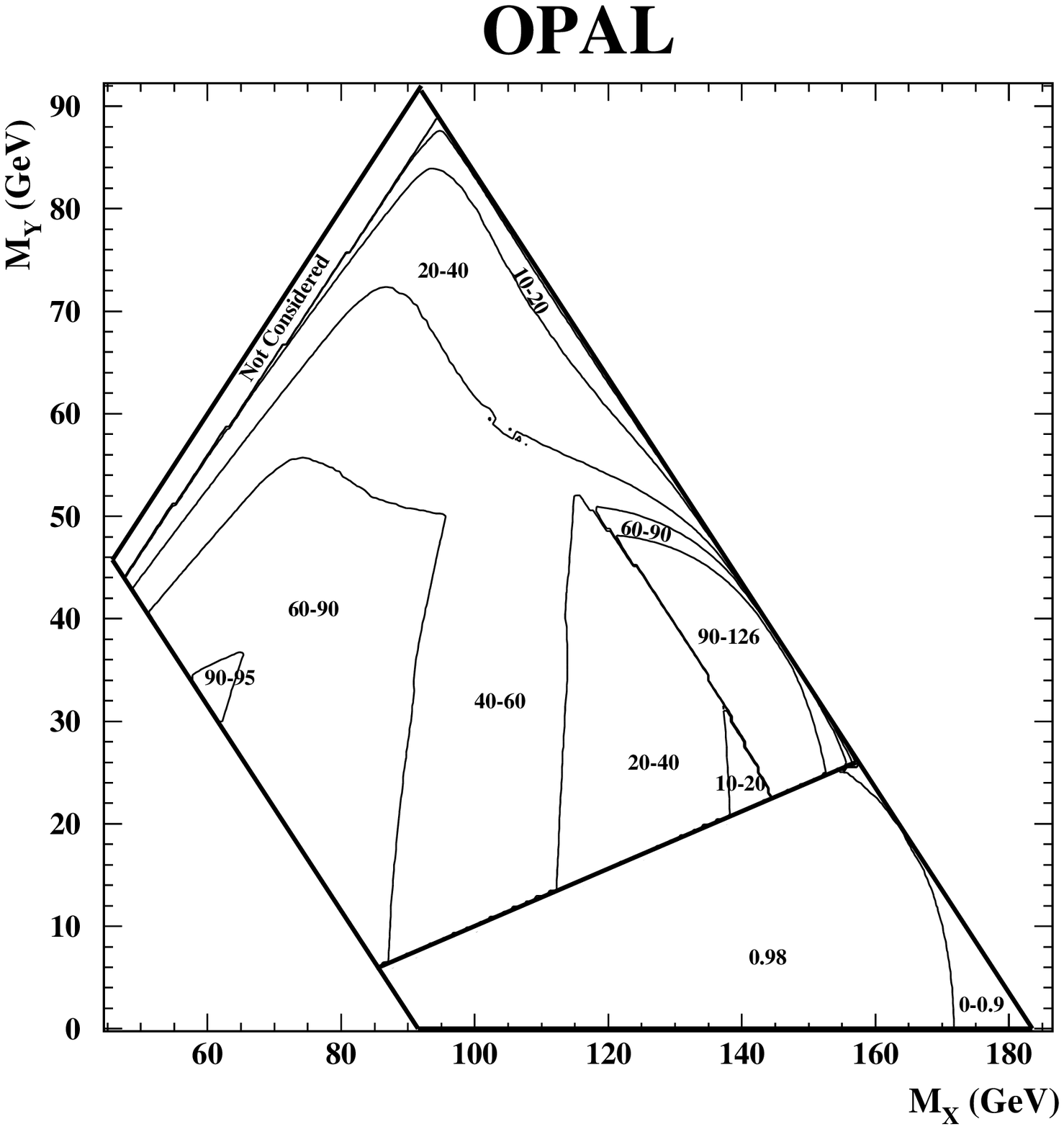}}
        \caption{
                Expected number of single-photon events from the
                process $\eetonnggbra$ at $\roots$=183 GeV which are
                consistent with each set of mass values ($\mx$, $\my$)
                for the process $\eetoXY$, $\XtoYg$.
                This figure gives the expected Standard Model 
                contribution to Figure~\ref{f:sp_evsXY_data}.
                KORALZ was used to model the $\eetonnggbra$ process.
                The boundaries and delineated regions are as defined for
                Figure~\ref{f:sp_XYeff}.}
        \label{f:sp_evsXY_mc}
\end{figure}

\begin{figure}[ht]
        \centerline{\epsffile{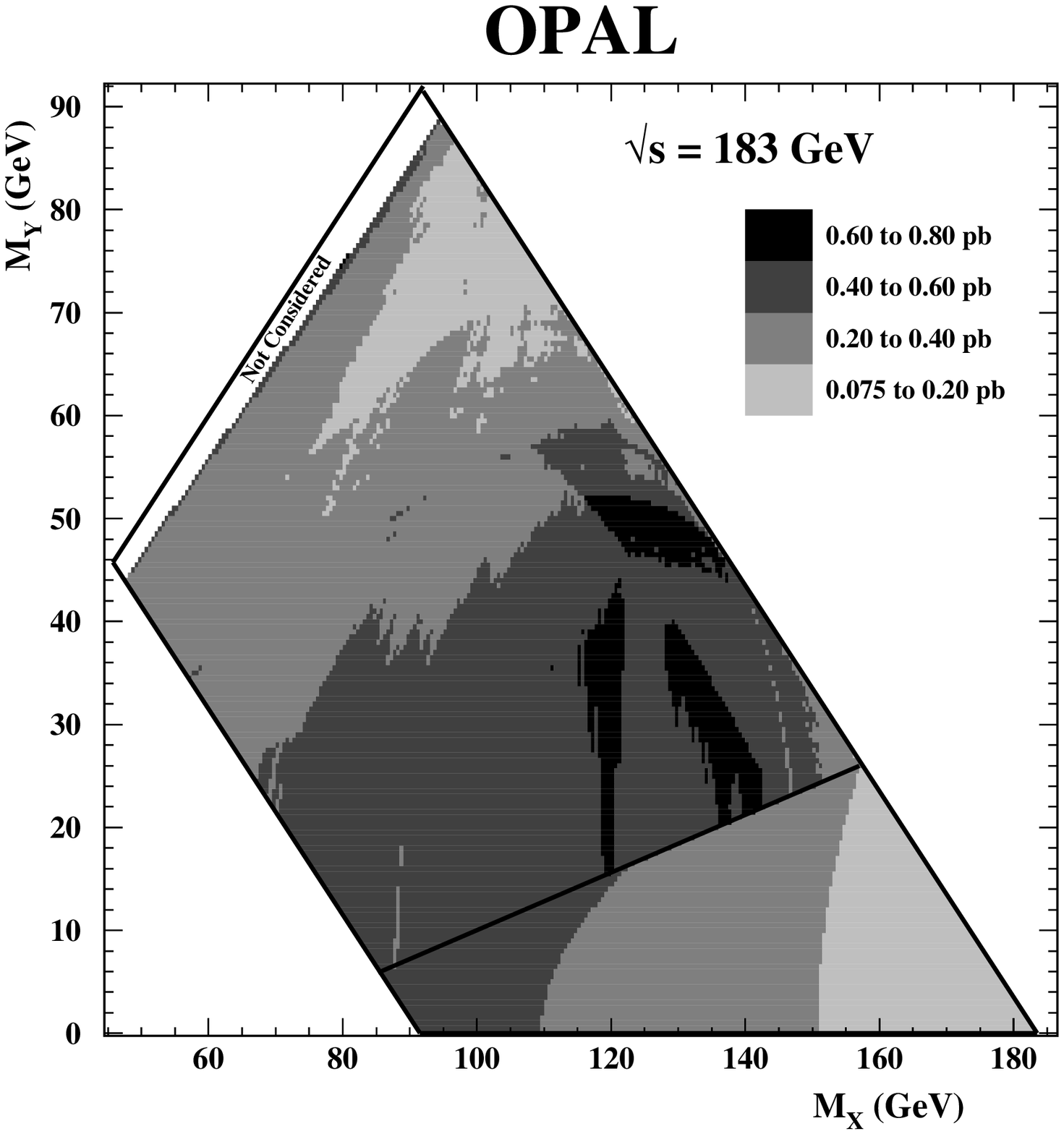}}
        \caption{The 95\% CL upper limit on $\sigbrXY$ 
                at $\roots = 183$ GeV
                as a function of $\mx$ and $\my$.
                The boundaries and delineated regions are as defined for 
                Figure~\ref{f:sp_XYeff}.}
        \label{f:sp_XYlim_all}
\end{figure}

The number of selected events in the data consistent with 
each $(\mx,\my)$ value is shown in Figure~\ref{f:sp_evsXY_data}
and can be compared with the number expected from Standard 
Model $\eetonnggbra$ 
events as shown in Figure~\ref{f:sp_evsXY_mc}.
In general there is good agreement,
and we proceed to set upper limits at 95\% confidence level (CL) 
on the cross-section times branching ratio, $\sigbrXY$, which are shown in
Figure~\ref{f:sp_XYlim_all}.
The upper limits are calculated taking into account the 
expected number of Standard Model $\eetonnggbra$ background events 
estimated from KORALZ using the method described in \cite{PDG96}.
Background from sources other than $\eetonnggbra$, including the estimated 
cosmic ray and beam related background, is
intentionally not taken into account in the limit calculations.
The resulting upper limits range from 0.075 pb to 0.80 pb.

The systematic error on the efficiency for selecting events from potential
new physics sources is due to the effects already discussed in
section 4.1, as well as
the uncertainty on the efficiency parametrization across the 
($\mx,\my$) plane.   The parametrization was compared to efficiencies
obtained from the fully simulated
Monte Carlo samples at 40 selected $(\mx, \my)$ points, and a
resulting systematic error of 1\% (absolute) was assigned across the 
plane. The total relative systematic error varies from 2 to 6\%
depending on $\mx,\my$; its effect on the upper limits is small, and is
calculated according to \cite{systerr}. 
Uncertainties in the $\nnggbra$ background estimate are also taken into
account.  Unlike the systematic error on the efficiency, a relatively 
small uncertainty in the background estimate can have a significant impact
on the resulting 95\% CL limit, especially when the number of expected
events is large.  To account for this uncertainty, a convolution is 
performed within the upper limit calculations.  Contributing sources to
the background uncertainty are:
the factors considered in the cross-section measurement (3.5\%), 
the estimated theoretical uncertainty in the $\nnggbra$ cross-section (2\%) 
based
in part on a comparison between the KORALZ and NUNUGPV98\cite{NUNUGPV98} event
generators, and a
contribution due to uncertainties in the energy scale.  This last 
contribution is dependent on the values of $\mx$ and $\my$; it is
calculated separately at each $(\mx, \my)$ point, and ranges from
negligible to 5\%.


\boldmath
\subsubsection{Search for $\eetoXY$, $\XtoYg$ ; Special case: $\myzero$}
\unboldmath
\label{sec:sp_results_my0}

\begin{figure}[ht]
        \centerline{\epsffile{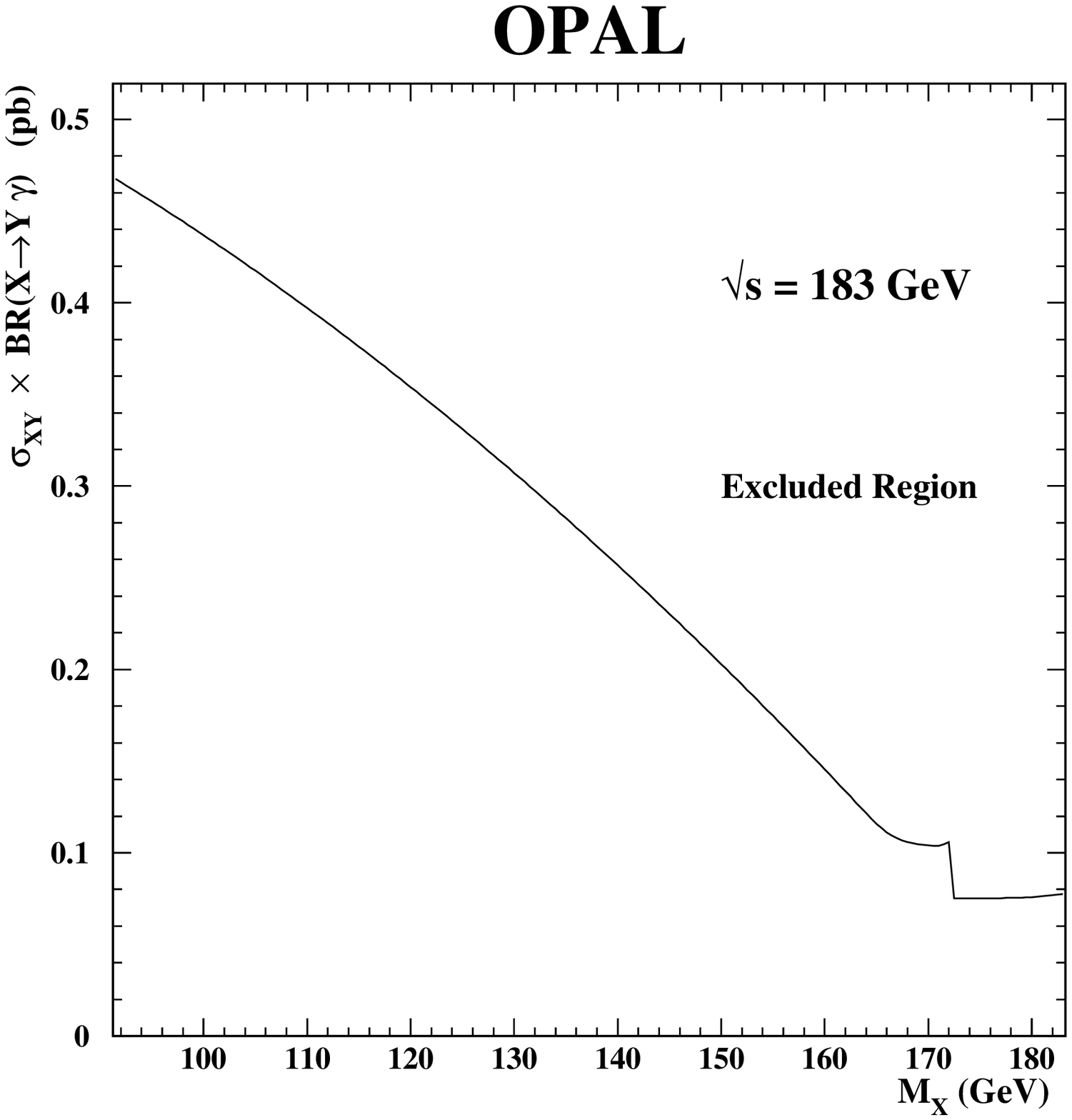}}
        \caption{The 95\% CL upper limits on $\sigbrXY$ at $\roots = 183$ GeV
                as a function of $\mx$, 
                assuming $\myzero$.}
        \label{f:sp_XYlim_massless}
\end{figure}

The case $\myzero$ is applicable to excited
neutrino models and to some supersymmetric models mentioned earlier.
The results presented above include this case and no separate
analysis is performed, but the results are highlighted here.
As described earlier, the 75 GeV recoil mass cut is applied 
for all $\mx$ hypotheses, so that the expected number of events is small. 
For example, in the range $91 < \mx \lesssim 170$~GeV, 
the expected contribution from $\nnggbra$ is $0.98\pm{0.12}$ events and
there is one event observed.  
Although the numbers of expected and observed events are constant in this range,
the efficiency of the recoil mass cut increases with increasing 
values of $\mx$, leading to decreasing values for the resulting
upper limits on $\sigbrXY$.
For $\mx \gtrsim 170$~GeV, the kinematic consistency 
requirements become more restrictive than the recoil mass cut.
There are no longer any events kinematically consistent with hypotheses of 
$\mx \gtrsim 173$~GeV; the background 
expectation in this region varies from $0.60 \pm 0.11$ events at 173 GeV 
to $0.04 \pm 0.01$ events at the kinematic limit.
The resulting upper limits on $\sigbrXY$ for $\myzero$ as a function of $\mx$
range from 0.46~pb to 0.075~pb,
as shown in Figure~\ref{f:sp_XYlim_massless}.


\boldmath
\subsection {Acoplanar-Photons}
\unboldmath
\label{sec:g2_results}

The acoplanar-photons selection applied to the 130, 136 and 183 GeV data
samples yields 2, 2 and 10 events, respectively, in good agreement
with the KORALZ predictions of $1.02\pm{0.02}$, $1.22\pm{0.02}$ and
$9.14\pm{0.09}$ events for the Standard Model $\eetonngggbra$ contribution. 
The expected contributions from other Standard Model processes and 
from cosmic ray and beam related backgrounds are small: less than 0.05, 0.05 and 0.1 
events, respectively.   
The numbers of events expected and observed at the three centre-of-mass energies 
are summarized in Table~\ref{tab:g2_nevt}. Also shown are the selection 
efficiencies for $\eetonngggbra$ events, within the kinematic acceptance of the selection,
and the corresponding cross-section measurements at each centre-of-mass energy. 
The OPAL measurements of the
cross-sections at $\roots =$ 130, 136, 161, 172 and 183 GeV are summarized in 
Table~\ref{tab:g2_xsec}. The measurements at 130 and 136 GeV are weighted averages of the results 
obtained from the 1997 and 1995 data samples. The latter results, as well as the results at
$\roots =$ 161 and 172 GeV, have been taken from our 
previous publication\cite{OPALSP172} and are corrected for 
the different definition of the kinematic acceptance.

Systematic errors arising from uncertainties on the electromagnetic calorimeter energy scale 
and resolution, the simulation of the detector material and consequent 
photon conversion probabilities, the 
integrated luminosity measurement and the detector occupancy estimate 
have been considered, and a relative systematic error of 8\% is assigned to the 
cross-section measurements. This comes dominantly from uncertainty on the energy
scale for low-energy photons and from comparison of different 
event generators\cite{OPALSP172}.

\begin{figure}[b]
\centerline{\epsfig{file=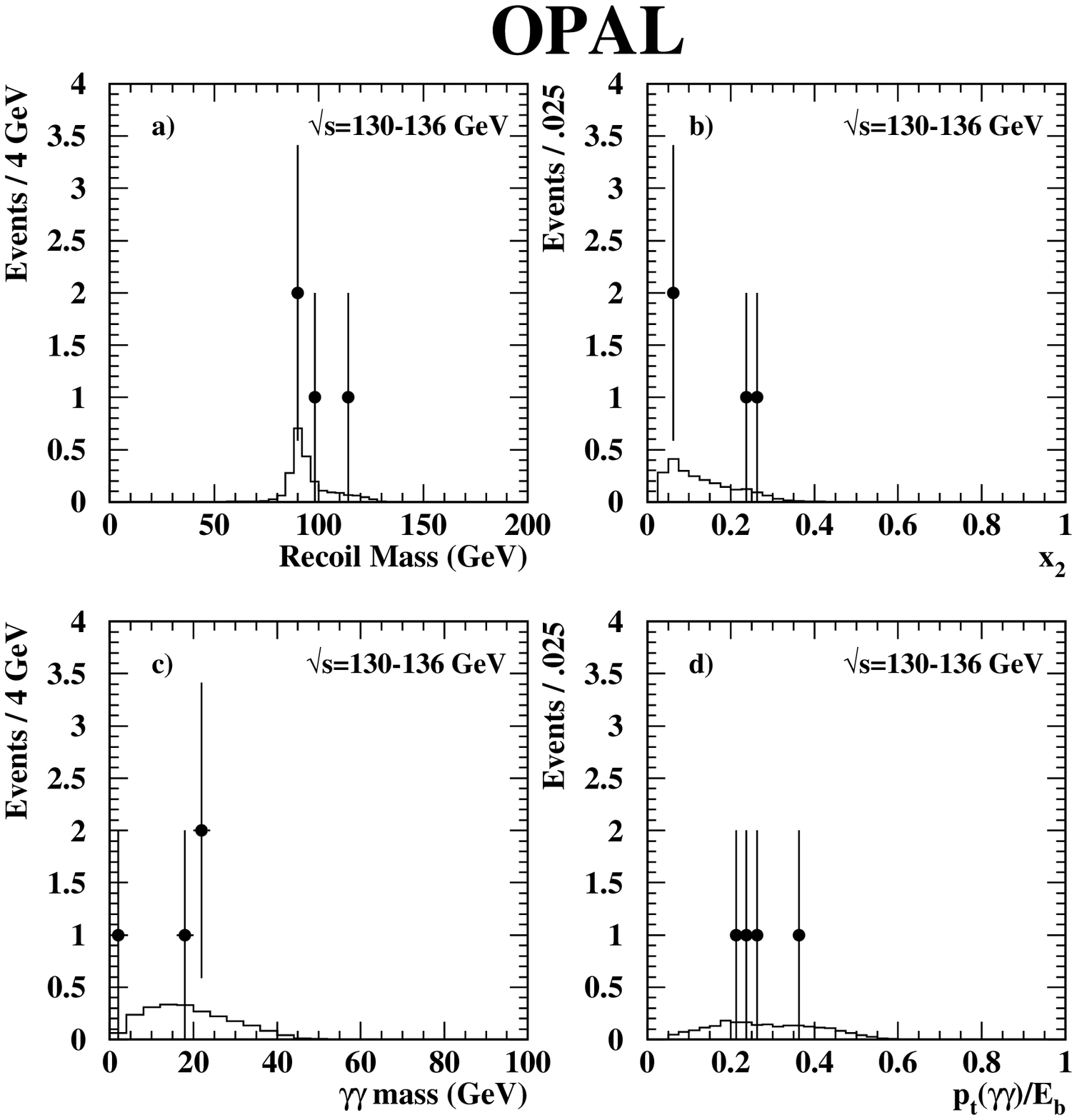,
width=16cm,bbllx=20pt,bblly=140pt,bburx=540pt,bbury=700pt}}
\caption{Plots of kinematic quantities for the
selected acoplanar-photons events from the
combined $\roots= 130-136$ GeV data sample.
a) Recoil mass distribution. b) Distribution of the scaled energy
of the second photon (x$_2$). c) Distribution of the invariant mass
of the $\gamgam$ system. d) Scaled transverse energy distribution for
the $\gamgam$ system.
The data points with error bars represent the selected OPAL data events.
In each case the histogram shows the expected contribution from 
$\eetonngggbra$ events, from KORALZ, normalized to the integrated luminosity
of the data. 
}
\label{f:g2_kin133}
\end{figure}

\begin{figure}[b]
\centerline{\epsfig{file=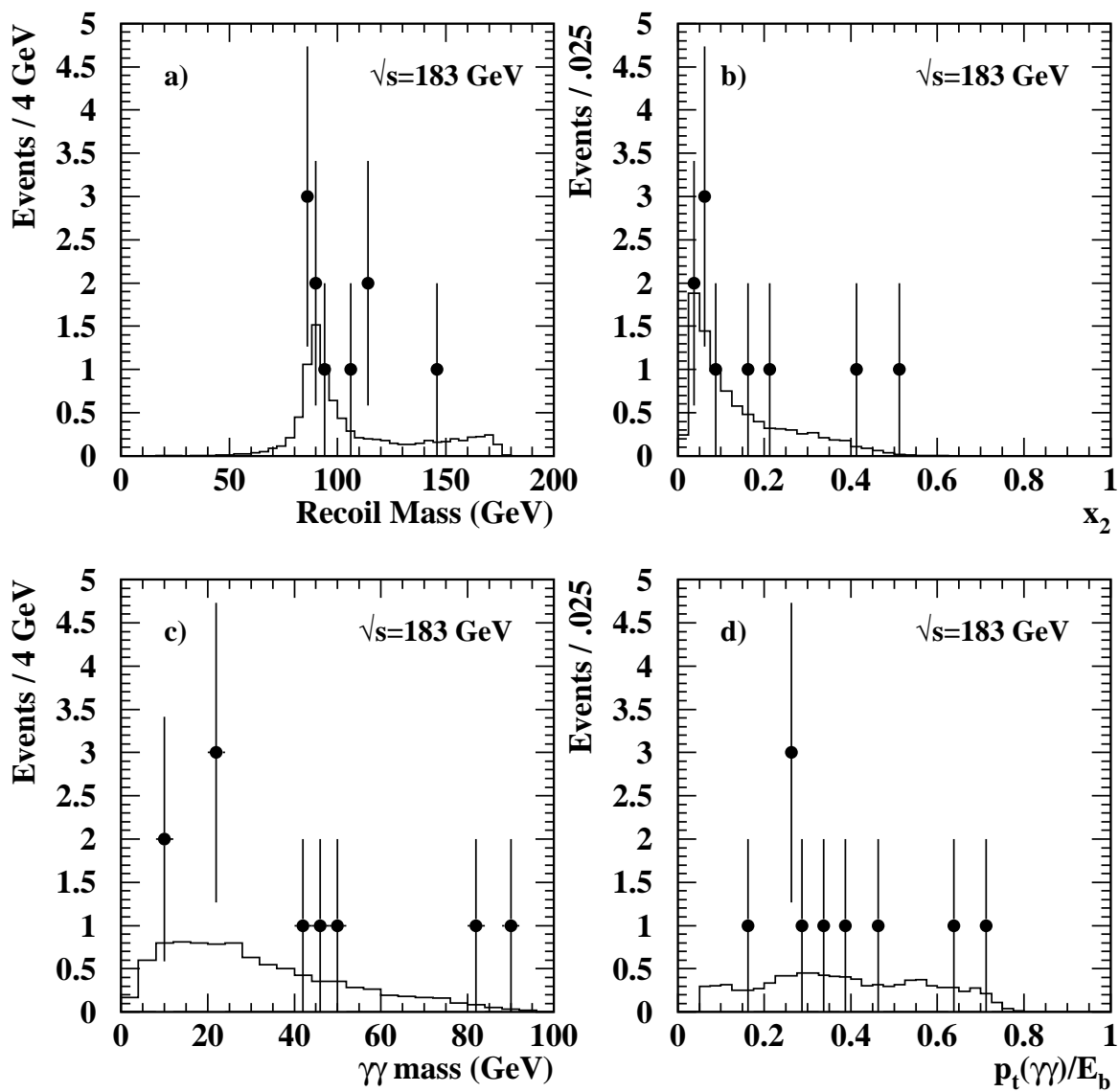,
width=16cm,bbllx=20pt,bblly=140pt,bburx=540pt,bbury=700pt}}
\caption{Plots of kinematic quantities for the
selected acoplanar-photons events for
$\sqrt{s}=$ 183 GeV.
a) Recoil mass distribution. b) Distribution of the scaled energy
of the second photon (x$_2$). c) Distribution of the invariant mass
of the $\gamgam$ system. d) Scaled transverse energy distribution for
the $\gamgam$ system.
The data points with error bars represent the selected OPAL data events.
In each case the histogram shows the expected contribution from 
$\eetonngggbra$ events, from KORALZ, normalized to the integrated luminosity
of the data. 
}
\label{f:g2_kin183}
\end{figure}

The kinematic properties of the selected events
in the combined 1997 data sample 
are displayed in 
Figures~\ref{f:g2_kin133} and \ref{f:g2_kin183}.
They are  compared with the predicted 
distributions for $\eetonngggbra$ obtained using the KORALZ generator normalized to the 
integrated luminosity of the data. In each case plot (a) shows the recoil mass distribution
of the selected acoplanar-photon pairs. These are peaked near the mass of the $\PZz$
as expected for contributions from $\eetonngggbra$. 
The resolution of the recoil mass is typically  2-4 GeV
for $M_{\rm recoil}\approx M_{\rm Z}$.
Plot (b) shows the distribution of the scaled energy of the least-energetic
photon. Plot (c) shows the $\gamma\gamma$ invariant mass distribution for which 
the mass resolution is typically 0.6-1.4 GeV. Plot (d)
shows the distribution in scaled transverse momentum of the 
selected two-photon 
system. The measured kinematic properties of the events are given in 
Table~\ref{tab:g2_prop}.
There were no selected events with three photons, 
compared to an expectation from KORALZ 
of $0.52\pm{0.02}$ events. 
  
In data taken in 1995 at $\roots =$ 130 and 136 GeV, 
we observed 8 events compared to $1.6\pm{0.1}$ expected from 
KORALZ\cite{OPALSP172}. However, the kinematic properties of these events, 
in particular the recoil mass distribution, agreed reasonably well with expectations.
It was demonstrated that plausible detector or beam related backgrounds 
do not contribute to this excess. Analysis of the 1996 OPAL data samples taken at 
$\roots =$ 161 and 172 GeV selected a number of events which was consistent
with the expected Standard Model background contributions.
In the data taken in 1997 at $\roots =$ 130 and 136 GeV, we select a total
of 4 events where $2.24\pm{0.03}$ are expected. The 1997 data are
therefore consistent with the KORALZ expectation.
However, the OPAL data samples at $\roots = 130$ and 136 GeV
continue to favour an excess of events over the expectation from KORALZ. 
Results on this topology have also been reported by the ALEPH collaboration
based on the analysis of 5.8 pb$^{-1}$ of data taken at $\roots = 130$ and 136 GeV in 1995.
That analysis has a kinematic acceptance similar to the one used by OPAL. 
No events were observed compared
to a Monte Carlo expectation of two events~\cite{ALEPH1P5}. 

As discussed in our previous publication, the status of
event generators and analytical calculations of the Standard Model process
$\eetonngggbra$ is not yet satisfactory.
We anticipate that on-going theoretical work by several authors
will result in increased understanding of 
the actual precision of current approaches, and lead to
improved approaches.
In particular, a new and more complete calculation
has recently appeared\cite{NUNUGPV98}.
For now, however, with contributions from higher order processes
not demonstrably under control, we 
do not know what theoretical uncertainty to assign to the 
KORALZ prediction.

In conclusion, the observed data excess over the KORALZ prediction 
at $\roots = 130$ and 136 GeV 
has not been resolved satisfactorily.  Analysis of all data collected 
at these centre-of-mass energies by the 
other LEP experiments could help to resolve whether 
the observed effect is real, rather than simply a statistical fluctuation
or a deficiency in the calculation of the Standard Model prediction.

In the data taken at $\roots = 183$ GeV, the agreement with expectations
from KORALZ is rather good. The only point of note is the selection of 
two events with a rather high invariant mass for the $\gamma\gamma$ pair. 
As seen in Figure~\ref{f:g2_kin183}c these events populate the high-mass 
tail of the distribution expected from $\eetonngggbra$. These events 
are both characterized by almost back-to-back photons
in the polar angle region above $\acosthe>0.9$.

\boldmath
\subsubsection{Search for $\eetoXX$, $\XtoYg$ ; General case: $\my\ge 0$}
\unboldmath
\label{sec:g2_results_allmy}

Selected events are classified as consistent with a given value of $\mx$ and $\my$
if the energy of each of the photons falls within the region kinematically accessible 
to photons from the process $\eetoXX$, $\XtoYg$, including resolution effects. 
The selection efficiencies at each generated grid point for the 
$\eetoXX$, $\XtoYg$ Monte Carlo events at $\roots = 183$ GeV are 
shown in Table~\ref{tab:g2_eff183}. 
These values include
the efficiency of the kinematic consistency requirement 
which is higher than 95\% at each generated point in the $(\mx,\my)$ plane.

\begin{figure}[b]
\centerline{\epsfig{file=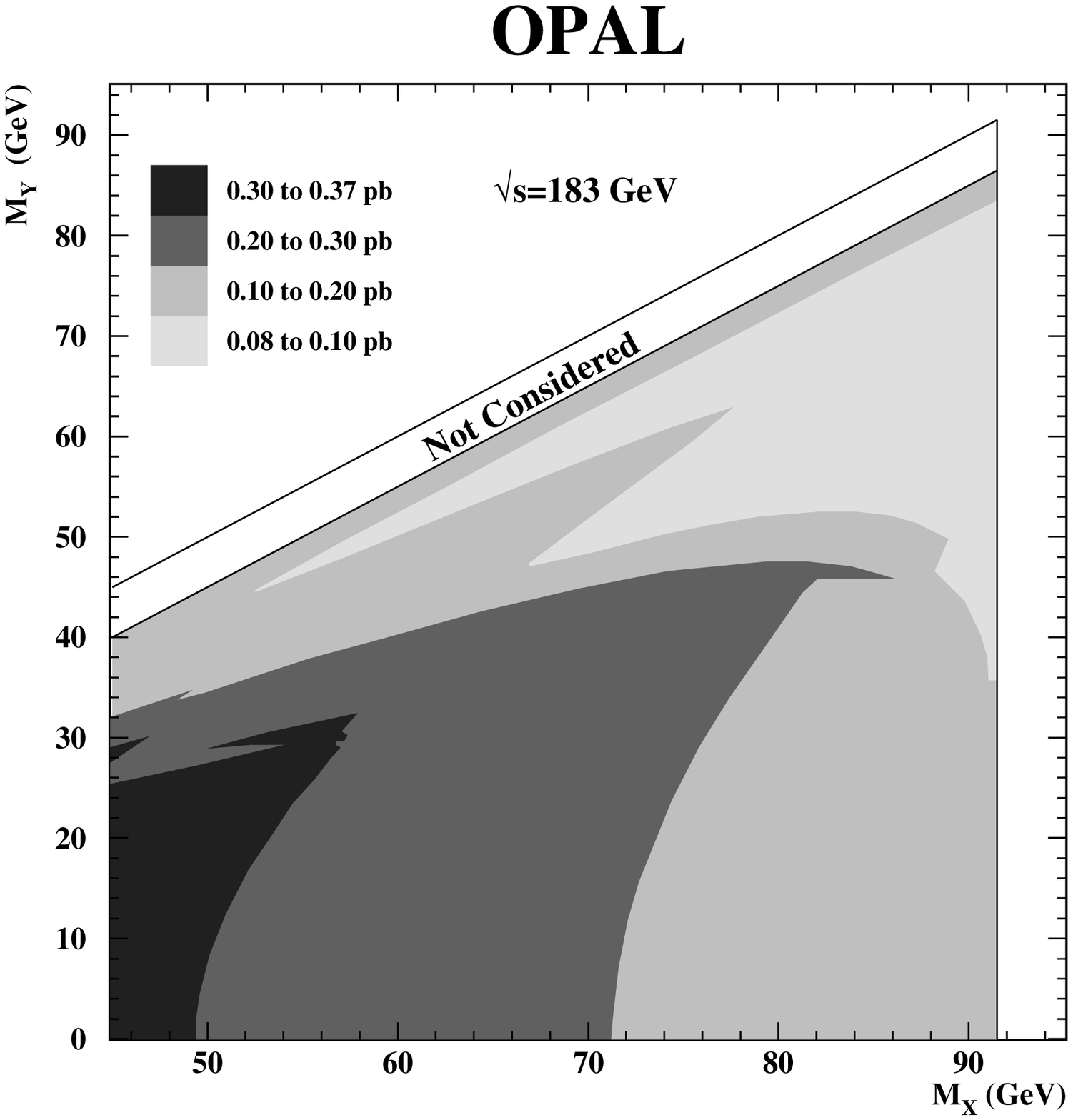,
width=16cm,bbllx=20pt,bblly=140pt,bburx=540pt,bbury=700pt}}
\caption{
The shaded areas show 95\% CL exclusion regions
for $\sigbrXX$  at $\roots = 183$ GeV. 
No limit is set for mass-difference values 
$\mx-\my < 5$ GeV, defined by the lower line above 
the shaded regions. The upper line is for $\mx=\my$.
}
\label{f:g2_mxmy}
\end{figure}

Figure~\ref{f:g2_mxmy} shows the 95\% CL exclusion regions for $\sigbrXX$.
The limits vary from 0.08 pb to 0.37 pb for $\mx >45$ GeV and $\mx - \my >5 $ GeV.
In the region 2.5 GeV $\le \mx -\my < 5.0$ GeV, the efficiency falls off rapidly
(see Table~\ref{tab:g2_eff183}).
However, even accounting for increased uncertainty on the efficiency, the limits
in that region are better than about 1 pb for all ($\mx$,$\my$).
Because of the uncertainties in the modelling of the Standard Model background,
as discussed earlier and in \cite{OPALSP172}, these limits and the limits presented below 
for this topology have been calculated without taking into account the background 
estimate. Events from $\eetonngggbra$ are typically characterized by a high-energy photon 
from the radiative return to the $\rm Z^0$ and a second lower energy photon. 
The kinematic consistency requirement is such that the two photons must have 
energies within the same (kinematically accessible)
region. As $\mx$ and $\my$ increase, the allowed range of energy for the photons 
narrows and fewer $\nngggbra$ events will be accepted. For the 10 selected events
at $\roots = 183$ GeV, the distribution of the number of events consistent with a given mass point 
($\mx$,$\my$) is consistent with the expectation from $\eetonngggbra$ Monte Carlo.

Systematic errors are due primarily to limited Monte Carlo statistics at the generated
($\mx,\my$) points and the uncertainty on the efficiency parametrization across the 
($\mx,\my$) plane. The combined relative uncertainty on the efficiency varies from about 
(3-6)\% across the plane (for $\mx - \my > 5$ GeV).
All systematic uncertainties are accounted for in the manner 
advocated in reference\cite{systerr}. This also applies to the limits for the $\myzero$ case, 
presented in the next section.

\boldmath
\subsubsection{Search for $\eetoXX$, $\XtoYg$ ; Special case: $\myzero$}
\unboldmath
\label{sec:g2_results_my0}

\begin{figure}[b]
\centerline{\epsfig{file=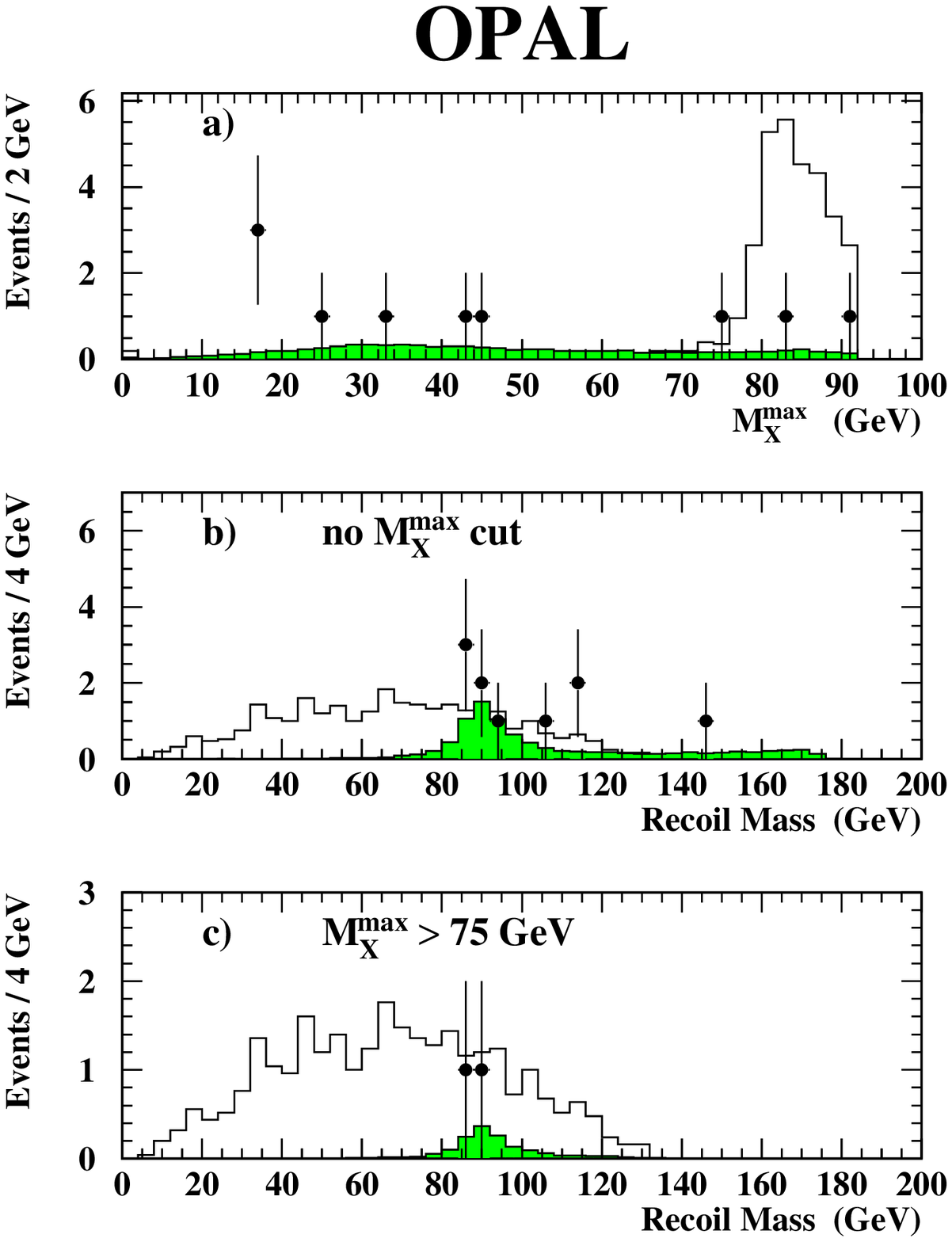,width=18cm,
bbllx=20pt,bblly=140pt,bburx=540pt,bbury=700pt}}
\caption{Properties of selected acoplanar-photons events at $\roots = $ 183 GeV.
a) The $\mxmax$ distribution.
b) The recoil mass 
distribution prior to any $\mxmax$ requirement. c) The same 
recoil mass distribution with a cut at $\mxmax > 75$ GeV 
(this cut is for consistency with an
$\mx$ of 80 GeV). In each plot the OPAL data are shown as points with error bars,
the shaded histogram shows the expected distribution for $\eetonngggbra$,
normalized to the integrated luminosity of the data, and the unshaded histogram 
shows the expected distribution for the signal process $\eetoXX$, $\XtoYg$ for 
$\mx = 80$ GeV. The three distributions for signal Monte Carlo 
are normalized to the same (arbitrary) production cross-section.}
\label{f:g2_mxmax_mrec}
\end{figure}

For the special case of $\myzero$ the kinematic consistency requirements differ
from those used for the general case.
One can calculate\cite{gravitinos2} the maximum mass,
$\mxmax$, which is consistent with the measured three-momenta
of the two photons, assuming a massless $\PY$.
A cut on $\mxmax$  
provides further suppression of the $\nngggbra$ background while retaining high 
efficiency for the  signal hypothesis. 
This is discussed in more detail in reference\cite{OPALSP172}.
We require
that the maximum kinematically allowed mass be greater than $\mx-5$ GeV,
which retains $97\pm{2}$\% relative efficiency for signal at all values 
of $\mx$ while suppressing much of the remaining $\nngggbra$ background. 
Figure~\ref{f:g2_mxmax_mrec}a shows the expected $\mxmax$ distribution for signal Monte
Carlo events with $\mx = 80$ GeV and for $\eetonngggbra$ Monte
Carlo events.  Also shown is the distribution of the selected data events.
In addition, we require that the recoil mass be less than 80 GeV,
which approximately maximizes the expected sensitivity of the analysis
for all $\mx$.
This cut retains more than 50\% of the signal efficiency, for
all values of $\mx$, and dramatically reduces the residual $\nngggbra$ background. 
This is most true in the 
region of low $\mx$ and remains valid up to the kinematic limit. 
A recoil mass cut is not applied in the massive Y case since this would lead to 
a large loss of efficiency in certain regions of the $(\mx,\my)$ plane. 
For the $\myzero$ case, the efficiencies calculated from Monte Carlo
events generated at 183 GeV are shown in Table~\ref{tab:g2_eff183_my0} 
after application of the event selection criteria and the cut on $\mxmax$, and then
after the additional requirement of $M_{\rm recoil}<80$ GeV.  Also shown in 
Figure~\ref{f:g2_mxmax_mrec} are the $M_{\rm recoil}$ distributions
for selected events with (b) no cut on $\mxmax$ and (c) 
$\mxmax > 75$ GeV (for consistency with $\mx = 80$ GeV).
No event survives the recoil mass cut; the expected number of Standard
Model events is $0.66\pm{0.03}$.
The expected number consistent with $\mx \ge 45$ GeV
is $0.36\pm{0.02}$ 
decreasing to 
$0.07\pm{0.01}$
expected events 
consistent with $\mx \ge 90$ GeV
as shown in Table~\ref{tab:g2_eff183_my0}.

\begin{figure}[b]
\centerline{\epsfig{file=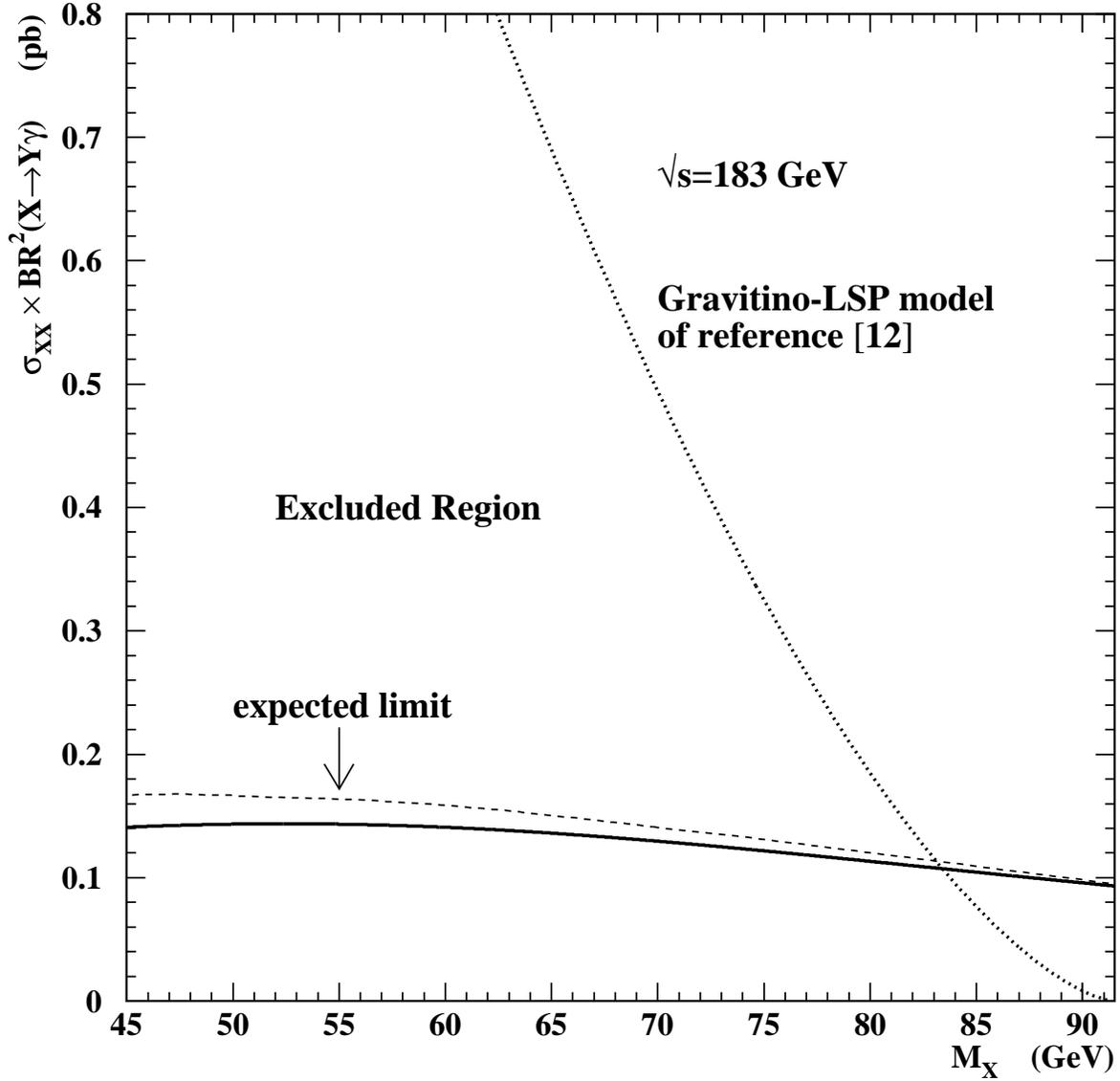,width=16cm,
bbllx=20pt,bblly=140pt,bburx=540pt,bbury=700pt}}
\caption{
95\% CL upper limit on $\sigbrXX$ 
for $\myzero$ (solid line). Also shown is the expected limit
(dashed line). The dotted line shows 
the cross-section prediction of
a specific light gravitino LSP model\cite{chang}. 
Within that model, $\lsp$ masses between 45 and 83 GeV are excluded 
at the 95\% CL. These limits assume that particle X decays promptly.}
\label{f:g2_limit_my0}
\end{figure}

Based on the efficiencies and the 
number of selected events, we calculate a 95\% CL upper 
limit on $\sigbrXX$ for $\myzero$ 
as a function of $\mx$. 
This is shown as the solid line in Figure~\ref{f:g2_limit_my0}.
The limit ranges from 0.094 to 0.14 pb for all $\mx$ from 45 GeV to the
kinematic limit.
Also shown as a dashed line is the expected limit, defined in
section~\ref{sec:sp_results_allmy}.
These limits can be used to set model-dependent limits on the mass of the 
lightest neutralino in supersymmetric models in which the 
NLSP is the lightest neutralino and the
LSP is a light gravitino ($\PX=\lsp, \PY=\Gravitino$).
Shown in Figure~\ref{f:g2_limit_my0}
as a dotted line is the (Born-level) cross-section prediction from
a specific light gravitino LSP model\cite{chang} in which
$M({\tilde{e}_R}) = 1.35M(\lsp)$, 
$M({\tilde{e}_L}) = 2M({\tilde{e}_R})$ and 
the  neutralino composition is purely gaugino (bino).
Within the framework of this model, $\lsp$ masses between 45 and 83 GeV
are excluded at 95\% CL. The expected
number of $\nngggbra$ background events consistent with $\mx \ge 83$ GeV
is $0.12\pm{0.01}$.

As described in section 2, the efficiencies over the full angular range have
been calculated using  isotropic angular distributions for production and 
decay of $\PX$. The validity of this model  has been examined
based on the angular distributions calculated for photino pair 
production in \cite{ELLHAG}. For models proposed in \cite{gravitinos}, the 
production angular distributions are more central and so this procedure is 
conservative. For a $1 + \cos^2{\theta}$ 
production angular distribution expected for t-channel exchange of a 
very heavy particle according to \cite{ELLHAG}, the 
relative efficiency reduction would be less than 2\% at all
points in the ($\mx,\my$) plane.
 

\section{Conclusions}
We have searched for photonic events with large missing energy in two 
topologies in data taken with the OPAL detector at LEP, 
at centre-of-mass energies of 130, 136 and 183 GeV.

%
%
In the single-photon selection, which requires at least one photon with
$x_{T} > 0.05 $ in the region $15^{\circ}<\theta<165^{\circ}$
($\acosthe < 0.966)$, a total of 21, 39 and 191 events 
are observed in the data for $\roots$~=~130, 136 and
183 GeV, respectively. These numbers are in agreement with the 
expectations of the 
KORALZ Monte Carlo generator for the Standard Model process $\eetonnggbra$.
The expected background is small.
We derive upper limits on the 
cross-section times branching ratio for the process  
$\eetoXY$, $\XtoYg$ for the general case of
massive $\PX$ and $\PY$. The limits vary from 0.075 to 0.80 pb in the
region of interest in the $(\mx,\my)$ plane and include 
the special case of $\myzero$, where the limit varies between
0.075 and 0.46 pb for the $\mx$ mass range from $M_{\rm Z}$
to 183 GeV.

The acoplanar-photons selection requires at least two photons with scaled energy
$x_{\gamma}>0.05$ within the polar angle region $15^{\circ}<{\theta}<165^{\circ}$
or at least two photons with energy $E_{\gamma}>1.75$ GeV with one satisfying
$\acosthe < 0.8$ and the other satisfying $15^{\circ}<{\theta}<165^{\circ}$.
In each case, the requirement 
$p_T^{\gamma\gamma}/E_{\rm beam} > 0.05$
is also applied. A total of 2, 2 and 10 events are selected from 
the data samples at
$\roots$ = 130, 136 and 183 GeV, respectively.
The KORALZ predictions for the contributions from 
$\eetonngggbra$ are, respectively, 1.02, 1.22 and 9.14 events. 
The number of events observed in the 1997 data samples
and their kinematic distributions are consistent
with expectations for $\eetonngggbra$. 
We derive 95\% CL upper limits on 
$\sigbrXX$ ranging from 0.08 to 0.37 pb for the general case of massive 
$\PX$ and $\PY$. 
For the special case of $\myzero$, 
the 95\% CL upper limits on $\sigbrXX$ range from 0.094 to 0.14 pb.
Due to the uncertainties in the current modelling of the Standard Model process
$\eetonngggbra$,  all limits from the acoplanar-photons analysis were
calculated without taking into account the background estimate.

For the single-photon and acoplanar-photons search topologies, 
the general case of massive $\PX$ and $\PY$ is relevant to supersymmetric models
in which $\PX = \nln$ and $\PY = \lsp$, with $\XtoYgs$ and $\lsp$ stable.
The special case of $\myzero$ is of particular interest for  
single and pair production of excited neutrinos and for
supersymmetric models in which the LSP is a light gravitino and the 
NLSP is $\lsp$ which decays as $\XtoYgg$.
For the latter scenario, the results of the acoplanar-photons search are used to place
model-dependent lower limits on the $\lsp$ mass.
A specific light gravitino LSP model\cite{chang} is excluded for the case
of promptly decaying neutralinos with masses between 45 and 83 GeV.

\section{Acknowledgements}

The authors wish to thank F. Piccinini for providing
cross-section results.

We particularly wish to thank the SL Division for the efficient operation
of the LEP accelerator at all energies
 and for their continuing close cooperation with
our experimental group.  We thank our colleagues from CEA, DAPNIA/SPP,
CE-Saclay for their efforts over the years on the time-of-flight and trigger
systems which we continue to use.  In addition to the support staff at our own
institutions we are pleased to acknowledge the  \\
Department of Energy, USA, \\
National Science Foundation, USA, \\
Particle Physics and Astronomy Research Council, UK, \\
Natural Sciences and Engineering Research Council, Canada, \\
Israel Science Foundation, administered by the Israel
Academy of Science and Humanities, \\
Minerva Gesellschaft, \\
Benoziyo Center for High Energy Physics,\\
Japanese Ministry of Education, Science and Culture (the
Monbusho) and a grant under the Monbusho International
Science Research Program,\\
Japanese Society for the Promotion of Science (JSPS),\\
German Israeli Bi-national Science Foundation (GIF), \\
Bundesministerium f\"ur Bildung, Wissenschaft,
Forschung und Technologie, Germany, \\
National Research Council of Canada, \\
Research Corporation, USA,\\
Hungarian Foundation for Scientific Research, OTKA T-016660, 
T023793 and OTKA F-023259.\\


\newpage
\begin{table}[b]
\centering
\begin{tabular}{|l||c|c|c|} \hline
Background process &
$\roots$=130 GeV & 
$\roots$=136 GeV &
$\roots$=183 GeV   \\ \hline \hline
$\eetomumu(\gamma)$ & .04 $\pm$ .01 & .05 $\pm$ .01 & .20 $\pm$ .04  \\ \hline 
$\ee \ra \ell^{+} \ell^{-} \nu\bar{\nu} (\gamma)$ & .02 $\pm$ .01 &
 .03 $\pm$ .01 & .12 $\pm$ .03  \\ \hline 
$\eetotautau(\gamma)$ & .01 $\pm$ .01 & .01 $\pm$ .01 & .03 $\pm$ .01  \\ \hline 
$\eetogg(\gamma)$ & $<$.02 & $<$.03 & .03 $\pm$ .02  \\ \hline \hline
Total physics bkgd. & .07 $\pm$ .02 & .09 $\pm$ .02 & .38 $\pm$ .05  \\ \hline 
Other bkgd. & .02 $\pm$ .02 & .02 $\pm$ .02 & .39 $\pm$ .14  \\ \hline \hline
Total background & .09 $\pm$ .03 & .11 $\pm$ .03 & .77 $\pm$ .15  \\ \hline 
\end{tabular}
\caption{Numbers of events expected from various
background processes contributing to the 
single-photon event sample
for the three centre-of-mass energies.
The Standard Model background contributions
are given by process.
Also shown are the expected other backgrounds coming 
from cosmic rays and beam related sources. 
The errors shown are statistical and the upper limits
are at 68\% CL.
}
\label{tab:sp_background}
\end{table}
%
\begin{table}[b]
\centering
\begin{tabular}{|c||c|c|c|c|c|} \hline
$\roots$(GeV) & 
$\mathrm N_{obs}$ &
${\mathrm N}_{\nnggbra}$ & 
$\mathrm N_{bkg}$ &
${\epsilon}_{\nnggbra}$(\%) & 
$\mathrm {\sigma}_{meas}^{\nnggbra}$(pb) 
 \\ \hline \hline
130 & 21 & 25.8 $\pm$ 0.1 &
  0.09 $\pm$ 0.03  & 77.0 $\pm$ 0.4 & 11.6 $\pm$ 2.5  \\ \hline 
136 &  39 & 31.2 $\pm$ 0.2 &
  0.11 $\pm$ 0.03  & 77.5 $\pm$ 0.4 & 14.9 $\pm$ 2.4  \\ \hline 
183 & 191 & 201.3 $\pm$ 0.7 & 
  0.77 $\pm$ 0.15  & 74.2 $\pm$ 0.3 & \phantom{1}4.71 $\pm$ 0.34 
 \\ \hline \cline{1-5}
\end{tabular}
\caption{For each centre-of-mass energy, the table shows the number of events
from the single-photon selection
observed in the 1997 OPAL data, the number expected based on the KORALZ 
$\eetonnggbra$ event generator and the number of events expected
from backgrounds. 
Also shown are the efficiencies for $\eetonnggbra$
within the kinematic acceptance of the single-photon selection
(defined in section 3)
and the background-subtracted measured cross-sections
within the kinematic acceptance. The errors shown are statistical.
}
\label{tab:sp_results}
\end{table}
%
\begin{table}
\centering
\begin{tabular}{|c||c|c|c|c|} \hline
$\roots$(GeV) & 
$\mathrm N_{obs}$ &
${\mathrm N}_{\nngggbra}$ & 
${\epsilon}_{\nngggbra}$(\%) & 
$\mathrm {\sigma}_{meas}^{\nngggbra}$(pb) 
 \\ \hline \hline
130 & 2 & 1.02 $\pm$ 0.02 &
 $69.3\pm{1.0}$ & $1.23\pm{0.87}$ \\ \hline 
136 &  2 & 1.22 $\pm$ 0.02 &
 $69.1\pm{0.7}$ & $0.86\pm{0.61}$ \\ \hline 
183 & 10 & 9.14 $\pm$ 0.09 & 
 $67.9\pm{0.4}$ & $0.27\pm{0.09}$
 \\ \hline 
\end{tabular}
\caption{For each centre-of-mass energy, the table shows the number of 
selected acoplanar-photons events in the 1997 OPAL data and  
the number expected based on the KORALZ $\eetonngggbra$ event generator. 
Also shown are the efficiencies for $\eetonngggbra$ events within the kinematic 
acceptance of the acoplanar-photons selection
(defined in section 3) and the corresponding cross-section measurements.
}
\label{tab:g2_nevt}
\end{table}

\begin{table}
\centering
\begin{tabular}{|c||c|c|} \hline
$\roots$(GeV) & $\mathrm {\sigma}_{meas}^{\nngggbra}$(pb) & 
$\mathrm {\sigma}_{KORALZ}^{\nngggbra}$(pb) \\ \hline \hline
130 & $1.49\pm{0.68}$  & $0.626\pm{0.010}$\\ \hline 
136 & $1.23\pm{0.56}$  & $0.526\pm{0.008}$ \\ \hline 
161 & $0.16\pm{0.16}$  & $0.330\pm{0.018}$ \\ \hline
172 & $0.32\pm{0.23}$  & $0.303\pm{0.017}$ \\ \hline
183 & $0.27\pm{0.09}$ & $0.247\pm{0.002}$ \\ \hline 
\end{tabular}
\caption{The measured cross-section for the process $\eetonngggbra$,
within the kinematic acceptance defined in section 3, for different
centre-of-mass energies. For $\roots =$ 130 and 136 GeV the measurements
are the weighted average of the results obtained from the 1997 data and
the results obtained from the 1995 data. The latter results, as well 
as those at $\roots =$ 161 and 172 GeV, are taken from our 
previous publication~[1] and have been corrected for the different 
definition of the kinematic acceptance.
The final column shows the cross-section predictions from KORALZ.
The quoted errors are statistical.
}
\label{tab:g2_xsec}
\end{table}

\begin{table}
\centering
\begin{tabular}{|c|c|c|c|c|c|c|c|c|c|c|}
\hline

$\sqrt{s}$ & $\rm x_1$ & $\rm x_2$ & $\rm cos\theta_1$ & $\rm cos\theta_2$ & $\phi_1$ & $\phi_2$ &
$M_{\rm recoil}$ & $\mxmax$ & $M_{\gamma\gamma}$ & $p_T^{\gamma\gamma}/E_{\rm beam}$ \\ \hline \hline 
130.0& 0.175&   0.059&   0.266&   0.264&    5.637&  5.006&  113.8&  29.8&   4.0 & 0.218 \\ \hline
130.0& 0.263&   0.266&   0.296&   0.956&    3.574&  2.162&   91.4&  54.3&  20.0 & 0.275 \\ \hline
135.9& 0.432&   0.063&  -0.801&   0.928&    3.037&  5.901&   99.1&   7.3&  22.2 & 0.236 \\ \hline
136.0& 0.367&   0.234&   0.482&   0.657&    2.372&  0.847&   88.9&  57.2&  22.7 & 0.373 \\ \hline \hline  
182.8& 0.484&   0.221&  -0.958&   0.125&    2.138&  1.318&  107.4&  74.6&  40.6 & 0.329 \\ \hline 
182.7& 0.674&   0.061&   0.867&   0.451&    4.367&  4.788&   94.7&  43.2&  11.8 & 0.387 \\ \hline 
182.7& 0.519&   0.518&   0.957&  -0.965&    4.848&  4.556&   84.1&  91.3&  91.2 & 0.282 \\ \hline 
182.7& 0.726&   0.096&   0.706&  -0.737&    5.076&  2.587&   90.3&  16.4&  47.1 & 0.464 \\ \hline 
182.7& 0.761&   0.026&   0.293&  -0.883&    4.921&  0.762&   87.2&  16.2&  22.4 & 0.721 \\ \hline 
182.7& 0.307&   0.059&   0.907&  -0.728&    2.103&  2.558&  146.9&  33.8&  20.6 & 0.166 \\ \hline 
182.7& 0.604&   0.028&   0.913&  -0.611&    0.051&  5.073&  112.6&  18.0&  20.3 & 0.254 \\ \hline 
182.6& 0.709&   0.067&   0.551&   0.844&    0.985&  1.005&   87.0&  45.5&   8.3 & 0.627 \\ \hline 
182.7& 0.541&   0.416&  -0.915&   0.901&    1.490&  6.115&   91.4&  82.7&  83.1 & 0.271 \\ \hline 
182.7& 0.541&   0.156&  -0.841&   0.903&    3.050&  5.363&  113.2&  25.3&  51.9 & 0.252 \\ \hline 
\end{tabular}
\caption[]{Kinematic properties of selected acoplanar-photons events.
Energies and masses are in GeV. Angles are in radians. 
The quantity $\mxmax$ is defined in 
section 4.2.2.}
\label{tab:g2_prop}
\end{table}

\begin{table}
\centering
\begin{tabular}{|c||c|c|c|c|c|}
\hline
$\mx$ & $\my$=0 & $\my=\mx /2$ & $\my=\mx-10$ & $\my=\mx-5$ & $\my=\mx-2.5$ \\ \hline \hline
90 & $71.8\pm{1.3}$ & $71.0\pm{1.4}$ & $65.5\pm{1.4}$ & $39.2\pm{1.5}$ & $6.8\pm{0.8}$ \\ \hline
85 & $72.5\pm{1.3}$ & $71.8\pm{1.3}$ & $64.4\pm{1.4}$ & $40.0\pm{1.5}$ & $7.6\pm{0.8}$ \\ \hline
80 & $71.9\pm{1.3}$ & $71.2\pm{1.4}$ & $65.7\pm{1.4}$ & $41.4\pm{1.5}$ & $5.9\pm{0.7}$ \\ \hline
70 & $71.5\pm{1.4}$ & $70.0\pm{1.4}$ & $63.9\pm{1.4}$ & $46.0\pm{1.5}$ & $6.6\pm{0.7}$ \\ \hline
60 & $73.1\pm{1.3}$ & $74.2\pm{1.3}$ & $64.3\pm{1.4}$ & $43.5\pm{1.5}$ & $7.7\pm{0.8}$ \\ \hline
50 & $73.7\pm{1.3}$ & $72.2\pm{1.3}$ & $64.3\pm{1.4}$ & $43.7\pm{1.5}$ & $9.4\pm{0.9}$ \\ \hline
\end{tabular}
\caption[]{Acoplanar-photons selection efficiencies (\%) 
for the process $\eetoXX$, $\XtoYg$ at $\roots = 183$ GeV for various
$\mx$ and $\my$ (in GeV). These values include the efficiency of
the kinematic consistency cuts. 
The efficiencies for the generated points at $\my = 20$ and $\my=\mx-15$
are not shown.
}
\label{tab:g2_eff183}
\end{table}

\begin{table}
\centering
\begin{tabular}{|c||c|c|c|}
\hline
 & Selection efficiency with & Selection efficiency with & $\rm N_{\nngggbra}$ \\
$\mx$ & $\mxmax>\mx-5$ GeV & $M_{\rm recoil}<80$ GeV & \\ \hline
\hline
90 & $71.3 \pm 1.4$  & $57.5 \pm 1.5$ & $0.07\pm{0.01}$ \\ \hline
85 & $71.6 \pm 1.4$  & $53.8 \pm 1.5$ & $0.10\pm{0.01}$ \\ \hline
80 & $70.0 \pm 1.4$  & $47.1 \pm 1.5$ & $0.13\pm{0.01}$ \\ \hline
70 & $69.2 \pm 1.4$  & $40.7 \pm 1.5$ & $0.18\pm{0.01}$ \\ \hline
60 & $70.4 \pm 1.4$  & $42.7 \pm 1.5$ & $0.25\pm{0.02}$ \\ \hline
50 & $70.4 \pm 1.4$  & $37.2 \pm 1.5$ & $0.31\pm{0.02}$ \\ \hline
\end{tabular}
\caption[]{
Acoplanar-photons event selection efficiency (\%), as a function of mass, 
for the process $\eetoXX$, $\XtoYg$, for $\myzero$ at 
$\roots = 183$ GeV. The first column shows the efficiency  
of the selection described in section~3.2, after the cut on $\mxmax$. 
The second column shows the efficiency (\%) after the 
additional requirement that $M_{\rm recoil} < 80$ GeV. The last column shows the 
expected number of events from the process $\eetonngggbra$ (KORALZ). 
The errors are statistical. 
}
\label{tab:g2_eff183_my0}
\end{table}

%
\newpage
%
\newpage
%
\newpage
%
\newpage
%
\newpage
%
%
\newpage
\newpage
%
\newpage
%
%
\newpage
%
\newpage

%
%
\clearpage
\newpage
\newpage
\newpage
%
%
%
\newpage
\newpage


%
\end{document}